\documentclass[12pt]{spieman}  

\usepackage{amsmath,amsfonts,amssymb}
\usepackage{graphicx}
\usepackage{setspace}
\usepackage{tocloft}
\usepackage{lineno}
\usepackage{soul}
\usepackage{xcolor}
\usepackage{float}
\usepackage{hyperref}

\title{Active focal-plane phase mask coronagraphy with a discrete pixelated device. I. Study of the theoretical performance trade-space}

\author[a,*]{Liurong Lin}
\author[b]{Axel Potier}
\author[a]{Ruben Tandon}
\author[a]{Lucas Marquis}
\author[a]{Jonas G. Kühn}
\affil[a]{University of Bern, Physics Institute, Division of Space Research and Planetary Sciences, Sidlerstrasse 5, Bern, Switzerland}
\affil[b]{LIRA, Université Paris Cité, Observatoire de Paris, Université PSL, Sorbonne Université, CNRS, F-92190 MEUDON, France }

\cftpagenumbersoff{figure}
\cftpagenumbersoff{table} 
\begin{document} 
\maketitle

\begin{abstract}Recent advances in high-contrast exoplanet imaging instrumentation have introduced the concept of adaptive coronagraphy. For example, liquid-crystal-on-silicon (LCoS) spatial light modulators (SLMs) can be used as programmable phase masks or digital micro-mirrors devices (DMDs) as configurable pupil apodizers. Adaptive coronagraphy offers the ability to adjust in real time to changing observing conditions and science goals, such as switching between blind surveys and follow-ups of known objects, or optimizing observations of multiple star systems including binaries and triples. At the same time, these active devices present challenges: finite spatial sampling, limited phase resolution, and the scalar nature of their modulation can all reduce coronagraphic performance. In this study, we look at the performance of a coronagraph utilizing pixelated discrete focal-plane masks (FPMs) in function of various key parameters notably: spatial sampling, phase resolution, temporal jitter, and calibration errors. The analysis includes several common FPM designs: vortex, four-quadrant phase mask (FQPM), Roddier $\&$ Roddier (R$\&$R), Dual zone phase mask (DZPM) and azimuthal cosine phase mask (ACM). Both monochromatic and  20 $\%$ broadband imaging conditions are considered, along with the absence or presence of a central obstruction from a secondary mirror in the telescope pupil, associated with variations in Lyot stop sizing. Our results provide insight into the error budget and key limiting parameters of scalar, pixelated FPM coronagraphs, for which contrast performance in ideal conditions is limited by spatial sampling and the chromaticity of the scalar phase modulation. Consequently, they are mainly relevant to ground-based high-contrast imaging, rather than to the substantially deeper contrast regimes required by future space-based observatories. While the analysis is currently mainly motivated by SLM-based implementations, the conclusions are expected to extend to other discrete pixelated focal-plane phase coronagraphs, including pattern-printed phase masks or next-generation light-modulation devices. 

\end{abstract}

\keywords{direct imaging of exoplanets, high contrast imaging, active coronagraphy, astronomical instrumentation, spatial light modulator}

{\noindent \footnotesize\textbf{*}Liurong Lin,  \linkable{liurong.lin@unibe.ch} }

\begin{spacing}{1.15}   

\section{Introduction}
\label{sect:intro}
Since the detection of 51 Pegasi b \cite{mayor1995jupiter}, the first exoplanet confirmed to orbit a Sun-like star, over 5,000 exoplanets have been discovered in recent decades using indirect methods such as radial velocity, transit light curves, and microlensing. In 2008, three exoplanets in the young star system HR 8799 were detected using direct imaging techniques \cite{marois2008direct}.  However, the challenges associated with direct imaging have resulted in a comparatively small number of exoplanets being found this way. Direct imaging faces two major challenges: the small projected angular separation between host stars and their planetary companions (typically sub-arcsecond for most systems), and the overwhelmingly large contrast between them. Contrast levels approaching $10^{-10}$ are required for the detection of Earth-like planets in reflected light around solar-type stars, which will represent the challenges that will be addressed by future space-based observatories. On the other hand, existing ground-based high-contrast imaging (HCI) facilities equipped with extreme adaptive optics, such as VLT/SPHERE\cite{SPHERE}, VLT/NACO \cite{NACO_1,NACO_2}, Gemini/GPI\cite{GPI}, Magellan telescope/MagAO-X\cite{MagAO-X} and Subaru/SCExAO\cite{SCExAO}, typically operate in contrast regimes of $10^{-3}$--$10^{-7}$, depending on wavelength, observing conditions, and wavefront control performance. These contrast levels are sufficient for the detection and characterization of young self-luminous giant exoplanets, and faint stellar companions \cite{chauvin2010deep,macintosh2015discovery, desidera2021sphere,langlois2021sphere, vigan2021sphere}. 

Within this context, coronagraphy is a key component of ground-based HCI instruments, enabling the suppression of on-axis starlight at small angular separations, originally developed by Bernard Lyot to study the faint emission of the Sun's corona \cite{lyot1939study}. 
Coronagraphs can be categorized into pupil-plane designs, focal-plane designs and hybrid designs such as apodized pupil lyot coronagraphs (APLC) \cite{APLC} and apodized vortex coronagraphs (AVC) \cite{AVC}. Focal-plane coronagraphs use amplitude or phase masks to block or diffract starlight at the focal plane. One of the first phase mask coronagraphs was the Roddier $\&$ Roddier (R$\&$R) \cite{roddier1997stellar}, which is sized so that the phase-shifted point-spread function (PSF) core contains roughly 50 $\%$ of the total stellar energy, enabling the phase-shifted and unshifted components to interfere destructively within the geometric pupil. However, its performance is limited by chromaticity and centering sensitivity. Building on this concept, the four quadrant phase mask (FQPM) \cite{rouan2000fqpm} was subsequently developed. It introduces a half-wavelength phase shift between adjacent quadrants, effectively canceling on-axis starlight while allowing off-axis companions to remain visible. However, its discontinuous phase transitions can introduce dead zones, reducing performance. To address these limitations, the vortex coronagraph was developed, employing continuous azimuthal phase ramps with even topological charges to eliminate dead zones and improve starlight suppression \cite{jenkins2008vortex}. Further advancements led to the Dual zone phase mask coronagraph, designed to mitigate R$\&$R chromaticity by introducing an additional phase shift in a secondary zone \cite{dual_zone}. More recently, the Azimuthal Cosine phase mask (ACM) has been proposed to further enhance broadband performance and robustness versus tip-tilt errors \cite{ACM}.


Early forms of adaptive coronagraphy were already explored in the early 2000s, for example with a variable-diameter occulting disk based on a compressed mercury drop \cite{Bourget2001}. In 2016, liquid-crystal-on-silicon (LCoS) spatial light modulators (SLMs) \cite{SLM_2, SLM_3} were first proposed as active, programmable focal-plane phase mask (FPM) coronagraphs \cite{kuhn2016digital}. With the high spatial resolution of SLMs, which often exceed 1 megapixel with pixel sizes in the order of 10 micrometers, SLMs can efficiently sample a telescope’s PSF. Optimized amplitude apodizers for phase-mask coronagraphs and arbitrary telescope apertures were proposed earlier as a way to mitigate diffraction from obstructed or complex pupils \cite{DMD_1}. More recently, digital micro-mirror devices (DMDs) have also been investigated as configurable pupil apodizers, enabling rapid reconfiguration of the pupil amplitude to adapt to different apertures or observing strategies \cite{DMDs}. 

Active and/or adaptive FPM coronagraphs can in principle allow the observer to choose the FPM design optimally as a function of the environmental conditions (seeing, residual tip/tilt jitter), the type of science targets (resolved or unresolved, single or multiple, blind survey or follow-up) and even to completely negate the impact of a non-ideal centrally-obscured telescope aperture \cite{oversize_LS_1}, which is increasingly relevant in the context of large segmented aperture with non-ideal merit functions. This flexibility allows for quickly implementing and prototyping any novel phase patterns outside the traditional ones, and this feature is particularly attractive for binary- or multiple-star systems, since the phase-mask pattern can in principle be tweaked to null several stars at the same time, and updated in real-time to follow the sky rotation of an altazimuth telescope mount, for the purpose of performing angular differential imaging (ADI). 

Despite the high potential, SLM-based active coronagraphs present clear weaknesses that may impact the performance in real-world scenarios.
The scalar phase delay of SLMs is highly chromatic by nature, negatively impacting performance in broadband light. \cite{SLM_4} 
The geometric-phase modulation has already been applied successfully in static coronagraphic elements such as vector vortex\cite{VVC_1} and vector apodizing phase plate coronagraphs\cite{VAPP}, using patterned liquid crystal masks. However geometric-phase SLMs are not yet available on the market, although some prototyping efforts have been reported \cite{geometric_SLM_1, geometric_SLM_2}. As a result, current adaptive coronagraph implementations still rely primarily on mature scalar SLM technology, despite its intrinsic chromatic limitations. In the longer term, the emergence of deformable mirrors with pixel pitches of a few tens of microns may provide an alternative route, as such devices could become suitable for focal-plane applications.
Additionally, SLMs require linearly polarized input light, reducing throughput by 50$\%$ for non-polarized light. Recent MEMS-based phase light modulators (PLMs), such as the Texas Instruments PLM and piston-type MEMS PLMs developed within the HEU ICT REALHOLO project\cite{realholo}, are potentially promising alternatives to liquid-crystal SLMs because they may provide fast, polarization-independent phase modulation \cite{PLM_1}. However, their present specifications, including stroke, phase resolution, wavelength range, fill factor, and optical quality, are not yet fully compatible with focal-plane coronagraphy \cite{PLM_2}. A detailed assessment of these technologies is therefore left for future work. The performance of the discrete pattern generated by the SLM remains an open topic, particularly near the central phase discontinuity of most FPM coronagraphs. 

To explore these aspects in a practical context, this work uses the instrument currently in commissioning: Programmable Liquid-crystal Active Coronagraphic Imager for the DAG telescope (PLACID)\cite{kuhn2016digital} as a representative implementation scenario, while the conclusions are intended to apply more generally to pixelated programmable phase-mask coronagraphy. After being installed and tested by the summer of 2026, PLACID will be the first high-contrast imaging instrument to field an LCoS SLM acting as a programmable active FPM coronagraph in H- (1.65 $\mu$m) and Ks-band (2.2 $\mu$m). The instrument serves as a pathfinder platform to test novel concepts of adaptive coronagraphy. These include coronagraphy of multiple-stars, self-calibration of non-common path aberrations (NCPAs) and Coherent Differential Imaging (CDI) in the time domain \cite{K_hn_2024}. 

This paper presents numerical simulations to assess the contrast, throughput, and signal to noise ratio (SNR) performance of commonly used non-ideal discretized FPMs, including the vortex, FQPM, R$\&$R, DZPM, and ACM. Simulations are conducted under both monochromatic and 20 $\%$ broadband conditions, representative of standard astronomical bandpass filters. The simulations also account for the presence or absence of a central obstruction caused by a secondary mirror in the entrance pupil, with or without its support structures. The goal is to understand the fundamental limitations of coronagraphy when pixelated devices are used as focal-plane masks in a realistic ground-based context, with the SLM-based implementation on PLACID serving as an illustrative example of a current existing system. The present study targets contrast regimes typical of ground-based instrument with adaptive optics, where instrumental effects, atmospheric residuals, and telescope pupil geometry dominate the performance budget.

\section{Method}

\begin{figure}[H]
    \centering
    \includegraphics[width=0.8\linewidth]{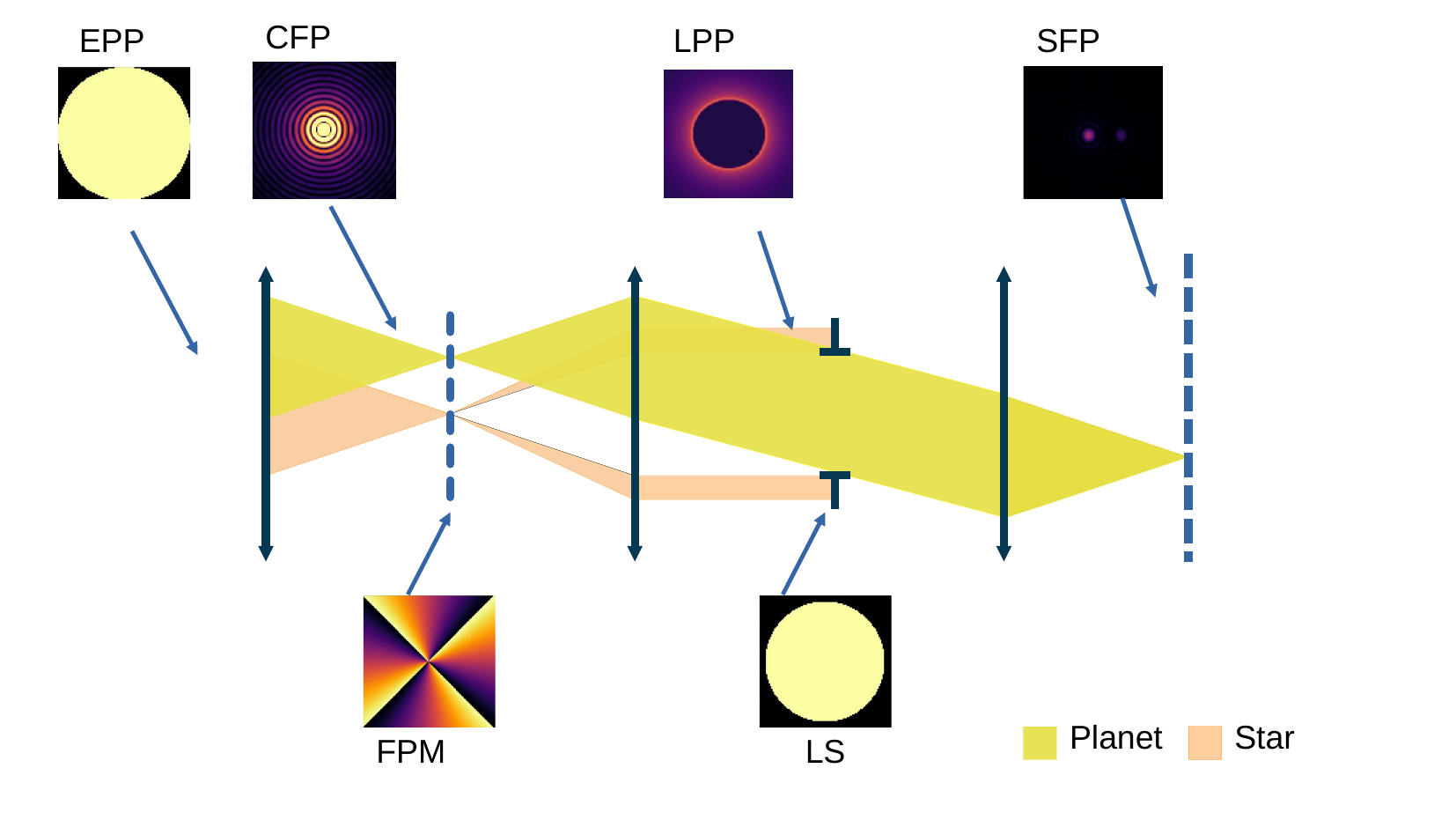}
    \caption{Cartoon illustration of light propagation through a coronagraph system. Key optical planes are labeled, including the entrance pupil plane (EPP), coronagraphic focal plane (CFP), Lyot pupil plane (LPP) and science focal plane (SFP). The image illustrates how star light and planet light propagate through each stage. The star light is diffracted and blocked in the LPP by the Lyot Stop (LS), while the planet signal passes through largely unaffected, leading to a high contrast image (HCI) at the SFP. Two key components FPM and LS are labeled at the bottom.}
    \label{fig:coro_scheme}
\end{figure}

The entire discrete coronagraphic system was simulated using Python, incorporating numerical methods for optical propagation. The supersampling function was adopted from the HCIPy library \cite{hcipy}, which initially simulated both the entrance pupil and Lyot post-coronagraphic pupil planes at a resolution of 800 $\times$ 800 pixels, before being downsampled to a final output of 100 $\times$ 100 pixels. This limited sampling of the pupil sets a contrast plateau, preventing the contrast from reaching the ideal level. This reduces aliasing artifacts that could lead to inaccuracies in the simulated diffraction patterns. The entrance pupil is sampled by a matrix with 100 $\times$ 100 pixels, balancing resolution with computational efficiency. Two-dimensional Fast Fourier Transform (FFT) is employed for optical propagation between the various pupils and focal planes to achieve comparable performance to Fresnel Transform but with higher computational efficiency \cite{FFT}. We use a zero-padded FFT propagation scheme because the pixelated focal-plane mask
is defined on a uniformly sampled grid with a fixed pixel scale. Although an Matrix Fourier Transform
could be evaluated on a uniform grid, its main advantage is flexible or localized sampling, which is not the primary requirement here. To optimize spatial sampling in the focal-planes, the 100 $\times$ 100 pixels entrance pupil is zero-padded in a computational grid of 10,000 $\times$ 10,000 pixels, providing up to 100 pixels per resolution element in the intermediate focal-planes. The schematic visualization of the simulated coronagraph system is shown in Figure \ref{fig:coro_scheme}.

\subsection{Spatial Sampling}
\label{sec:method_ss}
\begin{figure}[H]
\begin{center}
\includegraphics[height=0.8\linewidth]{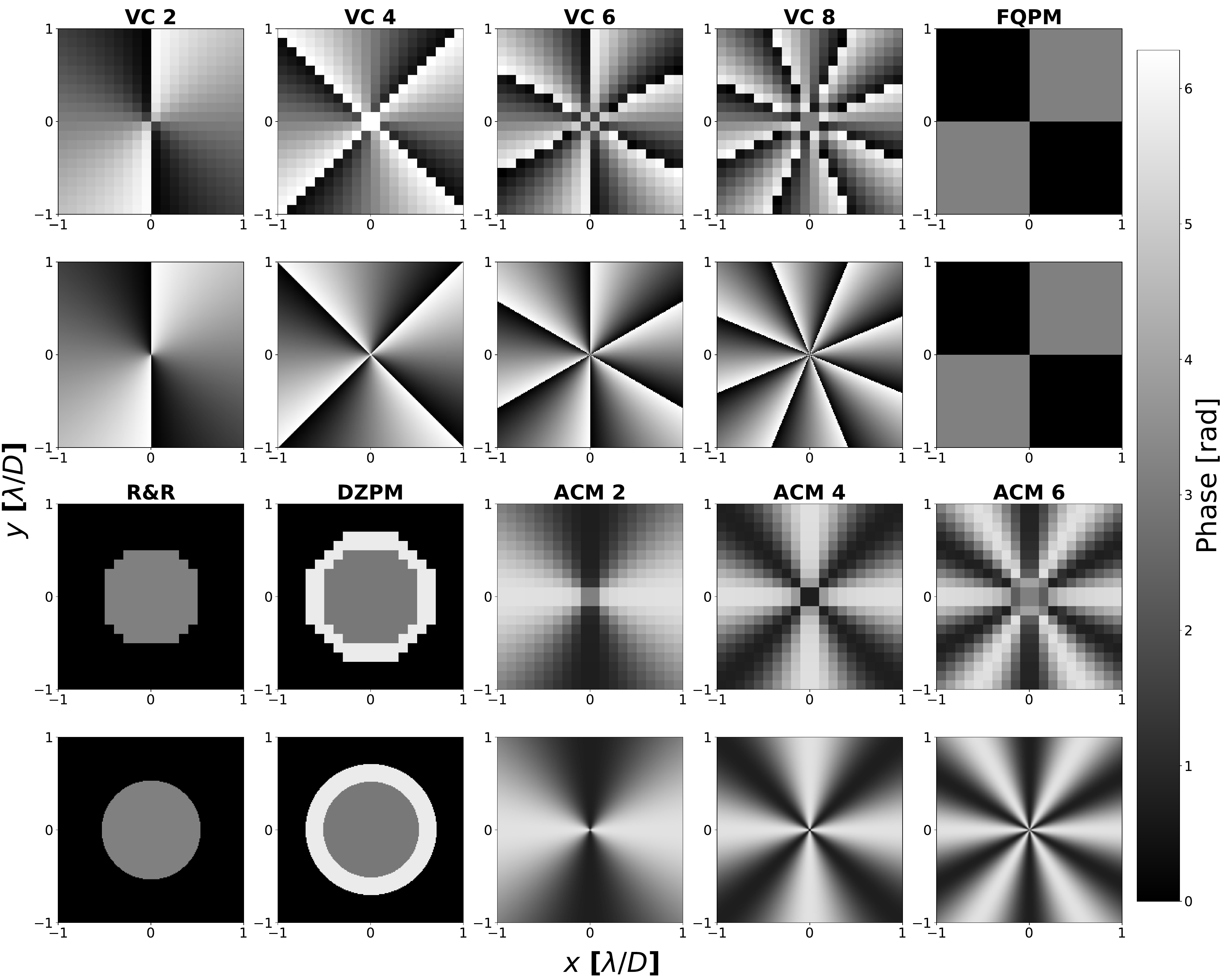} 
\end{center}
\caption
{ \label{fig:dicrete_mask}
Illustration of discrete FPMs with a spatial sampling of 10 pixels/($\lambda$/D) in \textbf{the first and third rows}, and ideal case with 100 pixels/($\lambda$/D) for comparison in \textbf{the second and fourth rows}. The figure displays the FPM region within $\pm$1 $\lambda$/D of the center. The center of the FPM is defined as the intersection of the four central numerical pixels.} 
\end{figure}

One of the intrinsic limitations of using discrete devices to implement FPM coronagraphs, such as LCoS SLMs, is the finite spatial sampling, which can impact the modulation of the incoming wavefront and the diffraction efficiency of the FPM pattern. Finite spatial sampling can also lead to a non-ideal or incomplete phase ramp, especially near the central singularity, which breaks the intended azimuthal phase symmetry of the mask and consequently degrades the coronagraphic null. Typically, most commercially available SLM panels count between 1 and 2 million pixels. The spatial sampling in the focal-plane is defined as the number of pixels per telescope resolution element ($\lambda$/D), where $\lambda$ is the wavelength and D is the telescope primary aperture diameter. In the simulation, a variety of FPM coronagraphic patterns commonly used or studied by the community (see Table \ref{tab:phase_mask}) are evaluated, with the following spatial sampling values: 2, 5, 10 and 100 pixels/($\lambda$/D) units in the coronagraphic focal-plane. The ``near ideal case" is defined as with 100 pixels/($\lambda$/D). Figure \ref{fig:dicrete_mask} illustrates the phase masks used in this work for the two spatial sampling rates. To decrease the spatial resolution of the FPMs, pixel binning is applied: as the sampling decreases, larger square regions are aggregated into single pixels, which effectively coarsens the phase pattern without altering the numerical extent of the focal plane array. This ensures consistent simulation conditions for all spatial samplings.
\begin{table}[h!]
\centering
\caption{Equations or parameters used by the FPMs in this work. (Note: $(x, y)$ denotes the focal-plane Cartesian coordinates with the center of the array as the origin. $\phi = \tan^{-1}\!\left(\frac{y}{x}\right)$ is the azimuthal angle in the focal plane.)}
\label{tab:phase_mask}
\begin{center}
\begin{tabular}{|l|p{11cm}|}
\hline
Phase Mask & Equation \\ \hline\hline

\rule[-1ex]{0pt}{3.5ex}
Vortex \cite{foo2005optical} & 
$\varphi(x,y) = l\phi$, where $l$ is the charge. \\ \hline

\rule[-1ex]{0pt}{3.5ex}
FQPM \cite{rouan2000fqpm} & 
$\displaystyle
\varphi(x,y) =
\begin{cases}
0, & \text{if } xy \ge 0, \\
\pi, & \text{if } xy < 0.
\end{cases}$ \\ \hline

\rule[-1ex]{0pt}{3.5ex}
R\&R \cite{roddier1997stellar} & 
$\displaystyle
\varphi(x,y) =
\begin{cases}
0, & \text{if } \sqrt{x^{2}+y^{2}} \ge 0.53\,\lambda/D, \\
\pi, & \text{if } \sqrt{x^{2}+y^{2}} < 0.53\,\lambda/D.
\end{cases}$ \\ \hline

\rule[-1ex]{0pt}{3.5ex}
DZPM \cite{dual_zone} & 
$\displaystyle
\varphi(x,y) =
\begin{cases}
0, & \text{if } \sqrt{x^{2}+y^{2}} \ge 0.705\,\lambda/D, \\
1.84\pi, & \text{if } 0.705\,\lambda/D > \sqrt{x^{2}+y^{2}} \ge 0.515\,\lambda/D, \\
0.94\pi, & \text{if } \sqrt{x^{2}+y^{2}} < 0.515\,\lambda/D.
\end{cases}$ \\ \hline

\rule[-1ex]{0pt}{3.5ex}
ACM \cite{ACM} & 
$\varphi(x,y) = z_{m}\cos(l\phi)$, where $l$ is the charge and $z_{m}$ is the $m$th zero of the type-J Bessel function. \\ \hline

\end{tabular}
\end{center}
\end{table}

To isolate the effects of spatial sampling, several simplifying assumptions are made in this initial batch of simulations. First, only monochromatic light is considered, eliminating chromatic scalar effects. Second, a perfect linear mapping between 0 and 2$\pi$  is assumed for an n-bit digital phase command, corresponding to a ``perfect phase calibration". Third, we assume zero wavefront errors (WFE) in the system, such as no  turbulence residuals from atmosphere or internal NCPAs. Finally, several 2nd-order SLM effects, including fill factor (gaps between pixels), phase jitter (temporal fluctuations in pixel phase response), crosstalk between adjacent pixels, and ghost reflections (unintended secondary reflections from the SLM surface) are not included at this point in the simulation. These factors, are deliberately omitted in this phase of study to provide a fundamental analysis of the role of FPM spatial sampling in the performance of coronagraphic systems, and will be considered at a later stage of our study.

\subsection{Phase quantization} 
A typical limitation of using a digital device, such as an SLM to program a coronagraphic FPM pattern is its limited phase resolution, incurred by the digitization over n numbers of bits. For the numerical simulation in python, phase values are typically represented using 64-bit floating-point precision, which we define as an ``ideal case". However, most commercially available LCoS SLM panels operate with only 8-bit or 10-bit phase resolution, corresponding to only 256 or 1024 discrete phase levels. This quantization introduces phase errors, that can affect the performance of the FPM coronagraph. The impact on coronagraphic performance will be studied, using the same assumptions as in Section \ref{sec:method_ss}.

\subsection{Broadband simulation}
\begin{figure}[H]
    \centering
    \includegraphics[width=0.8\linewidth]{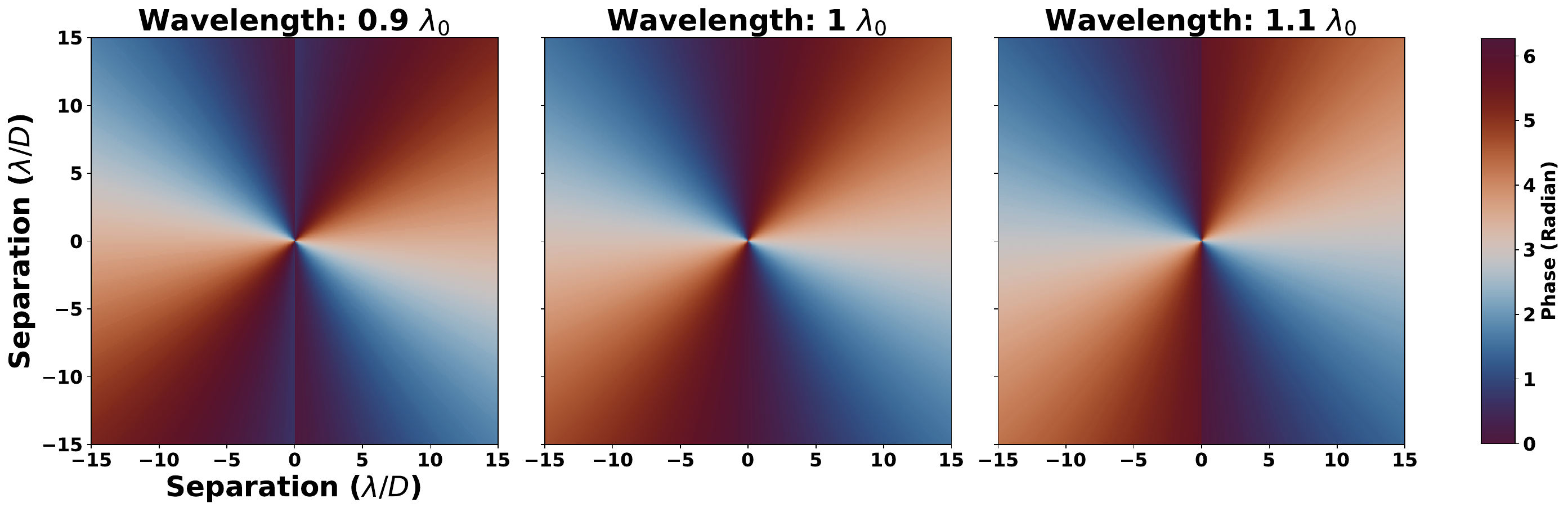}
    \caption{Computed phase shift induced by a vortex with charge 2 (VC2) for incident light at blue-shifted (0.9 $\lambda_{0}$) and red-shifted (1.1 $\lambda_{0}$) wavelengths.}
    \label{fig:phase_mask_wl}
\end{figure}

In order to simulate more realistic 20 $\%$ broadband light condition, representative of typical astronomical window bandpass filters, each broadband simulation includes five discrete  wavelength steps: 0.9 $\lambda_{0}$, 0.95 $\lambda_{0}$, $\lambda_{0}$, 1.05 $\lambda_{0}$ and 1.10 $\lambda_{0}$. The science focal plane (SFP) intensity maps from these five simulations are incoherently combined, with the following weight factors, in order to account for in-band contributions only: [0.125, 0.25, 0.25, 0.25, 0.125]. To reproduce the chromatic dependence of the diffraction limit in the focal plane, and to simulate the wavelength-dependent spatial sampling of the PSF with the coronagraphic FPM, the nominal entrance and Lyot pupil sizes (as defined for $\lambda_{0}$) are resized in inverse proportions of the wavelength \cite{PROPER}. Additionally, the phase shift values of the FPMs are modified based on the incoming wavelength according to Eq. \ref{eq:bb}:
\begin{equation}\label{eq:bb}
    \varphi_{\lambda} = \varphi_{0}\times\frac{\lambda_{0}}{\lambda}
\end{equation}
where $\varphi_\lambda$ is the actual phase shift imparted by the FPM for a given wavelength $\lambda$, $\varphi_{0}$ is the phase shift imparted by the phase mask at the central wavelength $\lambda_{0}$.
Figure \ref{fig:phase_mask_wl} illustrates the chromatic performance of a vortex phase mask with charge 2 under both blue-shifted (0.9 $\lambda_{0}$) and red-shifted (1.1 $\lambda_{0}$) incoming wavelength. In the blue shifted case, the phase shift imparted by the phase mask can exceed 2$\pi$ under 1$\pi$ of azimuthal rotation. Conversely, for the red-shifted light, the phase shift introduced by the mask falls short of 2$\pi$, resulting in an incomplete phase ramp.

\subsection{Telescope pupil parameters}
\label{sec:tele_para}
Both obstructed and unobstructed telescope configurations are considered in the simulation to evaluate the impact of a central obstruction on coronagraphic performance. The secondary mirror is modeled as 25$\%$ of the primary mirror diameter,  representative of the central obscuration
encountered in large ground-based telescope, such as the DAG telescope \cite{DAG_2022} and Palomar Hale 5-m telescope \cite{palomar}. It is well established that slight oversizing of the secondary obstruction in the Lyot stop (LS) can be beneficial to contrast for most FPMs, as the secondary mirror can diffract starlight into the dark region of the post-coronagraphic pupil \cite{oversize_LS_2, oversize_LS_1}.  However, this comes at the cost of reduced throughput for the off-axis source(s) of interest, which is highly detrimental on the SNR of any detected candidate. Hence there is a need for a metric to find the optimal balance between contrast and throughput. The SNR on a point source detection limited by stellar photon noise with coronagraphic HCI, can be expressed as Eq. \ref{eqn:SNR}.

\begin{equation}\label{eqn:SNR}
    \mathrm{SNR} = \frac{\eta_{p}}{\sqrt{\eta_{s}}}\epsilon\sqrt{N_{\star}}
\end{equation}
Here, $\eta_{p}$ is the fraction of remaining planet light, $\eta_{s}$ is the fraction of remaining star light, $\epsilon$ is the planet-to-star flux ratio and $N_{\star}$ is the total signal from the star \cite{SNR_eqn}. This formulation is convenient because it depends solely on the coronagraph parameters and the observed scene, rather than exposure time.
In Eq. \ref{eqn:SNR}, the factor $\xi = \frac{\eta_{p}}{\sqrt{\eta_{s}}}$ is isolated and used as a proxy for SNR to assess coronagraphic performance across various FPMs and LS sizing parameters. In this work, six different oversizing factors for the central obstruction are studied, with LS secondary diameter ratios of D$_{s}'$/D$_{s}$ = [1.0, 1.1, 1.2, 1.3, 1.4, 1.5], where D$_{s}'$ is the diameter of the LS obstruction and D$_{s}$ is the diameter of the entrance pupil obstruction. 
No undersizing factor of the primary mirror in the Lyot plane is included in our simulations, since this parameter is generally set by the pupil-centering stability (wobbling) budget of an instrument on a real telescope\cite{Kuhn_pupil}. Because this aspect lies beyond the scope of the present work, we restrict the present analysis to mitigation of the secondary-mirror obstruction, while keeping the outer Lyot stop diameter equal to that of the entrance pupil.
In addition, we also neglect the influence of the secondary support structure (spiders) in this study, as those were confirmed to be 2nd-order contrast offenders as compared to the parameters investigated here (see Appendix Section \ref{sec:ap_lpp_im}). 


\subsection{Temporal phase jitter}
\label{sec:tem_phase_jitter}
Most active digital devices that may be used as a programmable FPM coronagraph are subjected to some kind of temporal phase noise, or jitter. For an LCoS SLM, manufacturers characterize this behavior as phase ripple. It is defined as the ratio between the variation in intensity of the first order diffracted spot to the mean intensity, when a blazed phase grating is applied to the SLM. In the case of SLM panels manufactured by Meadowlark, as in use on the DAG/PLACID instrument, the phase ripple is specified to be less than 0.4$\%$ of one wave \cite{meadowlarkSLM2023}. Conservative estimates of 0.5 $\%$ and 5 $\%$ of 2$\pi$ root-mean-square (RMS) phase jitter on each digital FPM pixel are adopted in this work to study the impact of temporal phase noise on contrast performance. Assuming homogeneous pixel behavior over the panel, we simulate this effect by adding a random phase error as offset from the ideal value to each pixel. This error is drawn from a Gaussian phase distribution centered at the ideal value, with a standard deviation of $\sigma$ = 0.005 $\times$ 2$\pi$  $\approx$ 0.0314 rad and $\sigma$ = 0.05 $\times$ 2$\pi$  $\approx$ 0.314 rad respectively. We draw 20 FPMs and average the results.

\subsection{Calibration error}
\begin{figure}[H]
    \centering
    \includegraphics[width=0.8\linewidth]{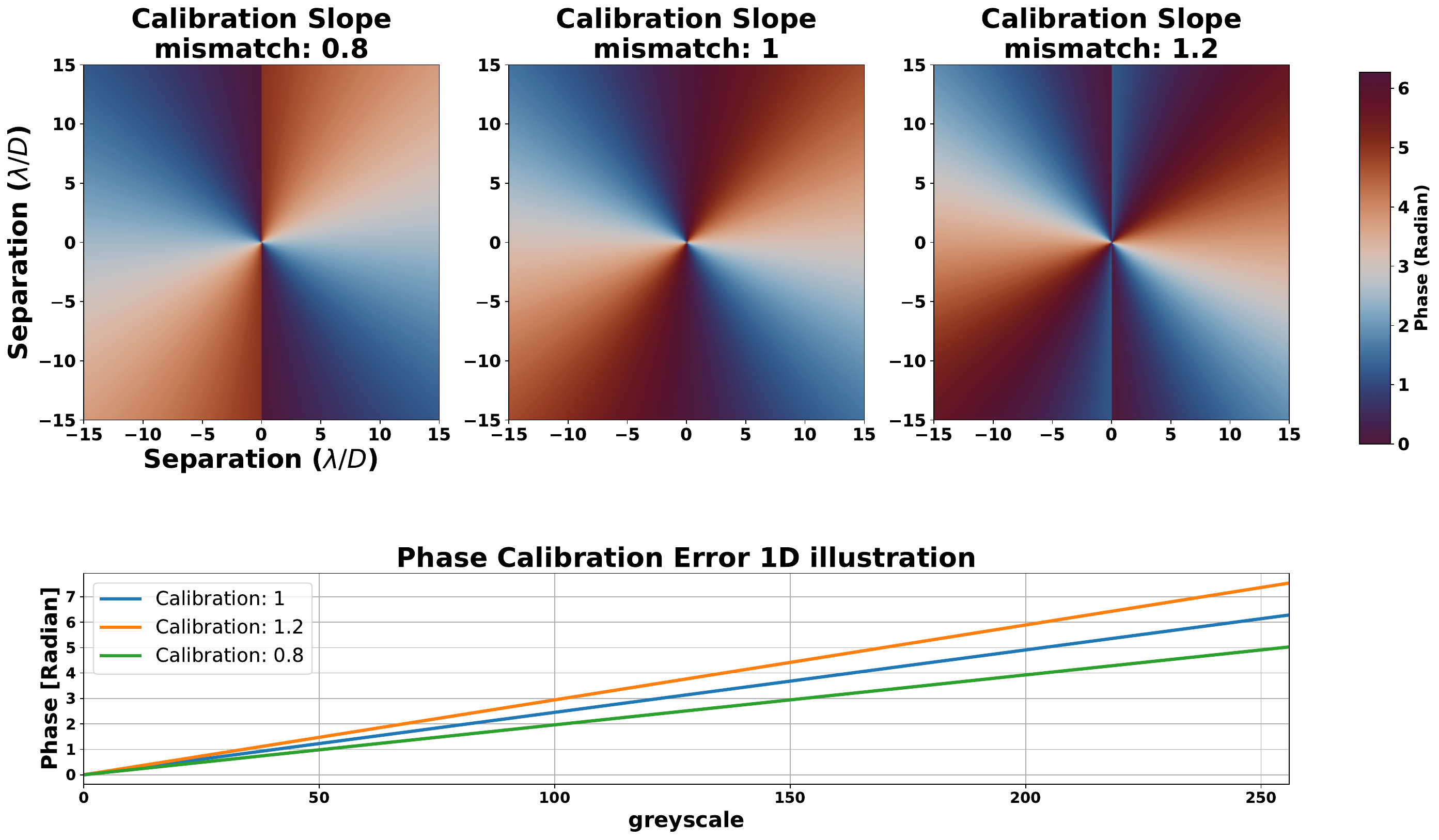}
    \caption{Computed phase shift induced by a VC2 for calibration error of 0.8 and 1.2, representing large phase calibration under- and over-shooting.}
    \label{fig:calibration_error}
\end{figure}

Digital devices like SLMs, but also static devices such as laser-printed patterns, are potentially subject to phase calibration errors. These errors typically arise from an imperfect mapping of the n-bit phase command (e.g. the 256 greyscale levels of the 8-bit SLM) over the 2$\pi$ optical phase. Assuming perfect linearity of the phase response (e.g. only the slope deviates from the ideal case), we simulate a range of phase slopes that result in imperfect mapping of the phase command over the ideal 2$\pi$ range. These deviations can introduce unwanted phase distortions, as illustrated in Figure \ref{fig:calibration_error} for Vortex with charge 2 (VC2). In both situations of over-shooting and incomplete phase ramp, the intended azimuthal phase symmetry of the vortex is no longer reproduced perfectly. This is expected to degrade the null depth of the on-axis stellar field and produce structured residual leakage in the SFP. FQPM and R$\&$R masks are especially sensitive to the exact phase value, because their operation relies on only two discrete phase states. Any deviation from the theoretical phase shift therefore produces residual stellar leakage and reduces the achievable null depth. The examples shown correspond to relatively large phase slope calibration errors (0.8, 1.0 and 1.2 slope changes).

\subsection{Fill factor}\label{sec:fill_factor}
The fill factor of an SLM refers to the fraction of each pixel's area that actively modulates the incoming wavefront, while the remaining area (typically pixel gaps or electrode edges) contributes no modulation and may scatter light.
The fill factor was experimentally measured to be 93 $\%$ in our lab. We utilize our simulation tool's ability to model up to 100 pixels/($\lambda$/D) sampling in the focal-plane. To simulate the effect of a non-unity fill factor, we artificially set the wavefront amplitude to zero along one row and one column within each 10 $\times$ 10 pixel block, assuming the FPM is sampled by a discrete device with a resolution of 10 pixels/($\lambda$/D). This scenario corresponds to about 81 $\%$ fill factor, which is a somewhat conservative estimate, as most commercially available SLMs offer a fill factor above 90 $\%$.


\subsection{Post-AO Residual Wavefront Error (WFE)}
\label{sec: method_atmosphere}
Simulated post-adaptive optics atmospheric wavefront errors residuals
are finally introduced, using the DAG/TROIA phase screen generator \cite{AO_simulation, AO_simulation_2}.
To this end, we assumed
a representative median observation scenario at H-band (1.6 $\mu m$), with a seeing of 1 arcsec, a natural guide
star (NGS) target magnitude of Vmag = 8 and a Zenith angle of 30$^\circ$ together with residual tip/tilt jitter levels of 5\% and 20\% of $\lambda/D$ rms \cite{AO_jitter}. The corresponding post-AO residual amplitudes, computed over the entrance pupil after piston removal and converted to wavefront-error RMS at the central wavelength (1.6 $\mu$m) is 89 nm RMS.
Each simulation was executed with 100 independently generated phase screens to account for time-average atmospheric turbulence over one exposure. An image cube containing 100 post-coronagraphic point spread functions (PSFs) is produced. The mean PSF was computed from the image cube to obtain the final averaged result, which represents the expected coronagraphic performance under the given observing conditions.
In the results presented below, these post-AO residual phase screens are introduced only in Section~3.7. Unless otherwise stated, the simulations discussed in the earlier subsections are performed without additional wavefront errors, in order to isolate the effect of the other parameters under study.
\subsection{Summary}
\label{sec:method_sum}
The results obtained using the methods described above are presented in Section \ref{sec:result_ss}, with the corresponding configurations summarized in Table \ref{tab:simulation_cases}.

\section{Results}
\label{sec:result_ss}
\begin{table}[ht]
\centering
\caption{Summary of simulation configurations used in this work.}
\label{tab:simulation_cases}
\begin{tabular}{|p{1.1cm}| p{2cm}|p{2cm}| p{2cm}| p{1.8cm}| p{1.5cm}| p{1.7cm}|}
\hline
Section & Bandwidth [\%] & Spatial Sampling [px/($\lambda$/D)] & Phase Resolution [bit] & Pupil Obstruction & Inner LS Oversize & AO Residuals \\
\hline
\ref{sec:spatial_sam} & $<1$      & 2--100 & 64    & No     & No  & No  \\
\hline
\ref{sec:met_gs} & $<1$      & 10     & 2--64 & No     & No  & No  \\
\hline
\ref{sec:met_obs} & $20$      & 10     & 8     & Yes/No & No  & No  \\
\hline
\ref{sec:met_bb} & $<1$ and $20$ & 10     & 8     & Yes/No & No  & No  \\
\hline
\ref{sec:met_ls} & $20$      & 10     & 8     & Yes    & Yes & No  \\
\hline
\ref{sec:met_pix_noise} & $<1$ and $20$ & 10     & 8     & Yes/No    & No  & No  \\
\hline
\ref{sec:met_ao} & $<1$      & 10     & 8     & Yes    & No  & Yes \\
\hline
\end{tabular}
\end{table}

\subsection{Impact of limited spatial sampling on the coronagraphic FPM contrast performance}
\label{sec:spatial_sam}
\begin{figure}[H]
    \centering
    \includegraphics[width=0.8\linewidth]{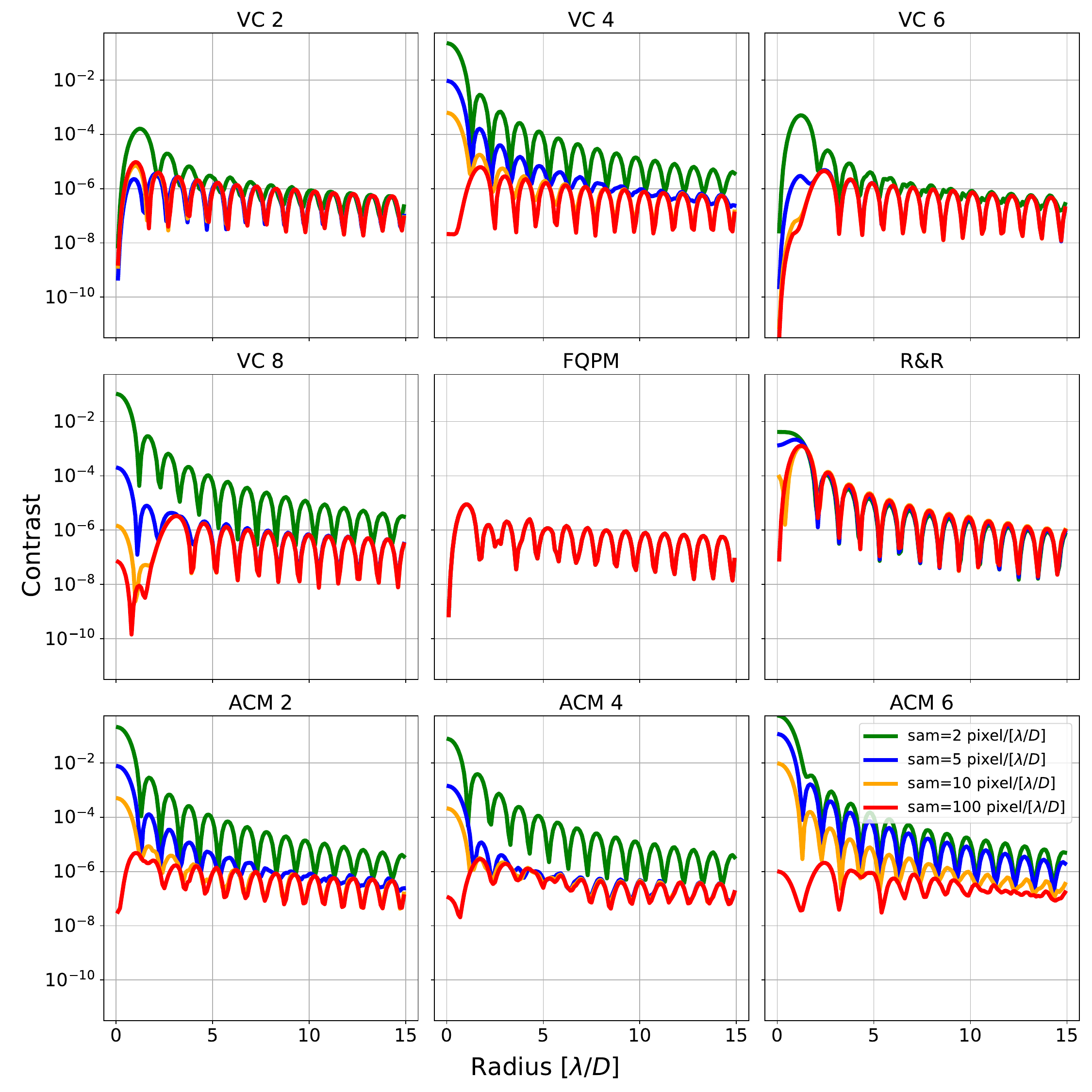}
    \caption{Azimuthally-averaged raw contrast curves in the SFP for discretized coronagraphic systems. Different FPM spatial sampling conditions are compared: 2, 5, 10, and 100 pixels/($\lambda$/D). Nine commonly used phase masks are evaluated. The analysis assumes monochromatic light, an unobstructed entrance pupil, 64-bit greyscale phase resolution for the FPMs and no wavefront error.}
    \label{fig:res_spatial_sam}
\end{figure}

We first consider the idealized scenario with monochromatic light, an unobscured telescope pupil and absence of wavefront errors. Figure \ref{fig:res_spatial_sam} illustrates the impact of spatial sampling on raw contrast in the post-coronagraphic science focal plane (SFP) for a range of FPM designs. Raw contrast here corresponds to the azimuthally averaged coronagraphic intensity, at a given angular separation from the on-axis star in the SFP, normalized by the peak intensity of the corresponding non-coronagraphic stellar PSF (with identical entrance pupil and Lyot stop)\cite{raw_contrast}. No post-processing, nor throughput correction for an off-axis source, is included in this metric, which can be interpreted as ``residual intensity” at a given angular separation. As an example, 10 pixels/($\lambda$/D) is typically the configuration currently implemented with the SLM on the DAG/PLACID instrument. As the spatial sampling is varied incrementally from 2 to 100 pixels/($\lambda$/D), all FPMs deliver improved starlight suppression.
Here, the case with 100 pixels per $\lambda/D$ is used as a near-ideal numerical reference for a pixelated FPM, rather than as the analytical limit of an ideal continuous phase mask, which would be unrealistic and would converge to a contrast level of zero in absence of wavefront error terms. However, even at 100 pixels per $\lambda/D$, the central singularity is then not perfectly resolved, since the four numerical pixels surrounding the optical axis necessarily approximate the phase discontinuity and converge towards a FQPM or a flat pattern, depending on the topological charge. In addition, the limited sampling of the entrance pupil, together with the absence of undersizing of the outer Lyot stop, prevents the contrast from reaching the ideal level.
The impact of limited spatial sampling is especially noticeable at small angular separation, where the aliasing effects are the most pronounced. Vortex masks and ACMs are the most sensitive to restricted spatial sampling, with the former especially affected for topographic charges that are multiples of four. For spatial sampling of at least 5 pixels/($\lambda$/D), the contrast performance of most FPMs beyond an angular separation of approximately 3 $\lambda$/D converges closely to that of the near-ideal case with 100 pixels/$(\lambda$/D). The only major exception is the ACM 6 pattern, which requires significantly finer sampling to mitigate substantial phase leakage in its central region. On the other hand, FQPM, and R$\&$R, to a lesser extend, are remarkably unaffected by limited spatial sampling even down to very low values. Both these two masks are insensitive to coarse sampling because they are binary phase masks with simple geometry and cancellation based on symmetry. Overall, FPMs with rapid azimuthal phase variation such as vortex and ACM require higher sampling rates to converge to their theoretical contrast capabilities. In addition to the nominal spatial sampling, the relative centering of the mask on the discrete pixel grid can also considerably influence the coronagraphic performance, particularly for masks converging towards a central singularity or a flat zero region. A dedicated analysis of this effect is presented in Appendix Section \ref{sec:ap_sub_pixel}.

\begin{figure}[H]
    \centering
    \includegraphics[width=0.5\linewidth]{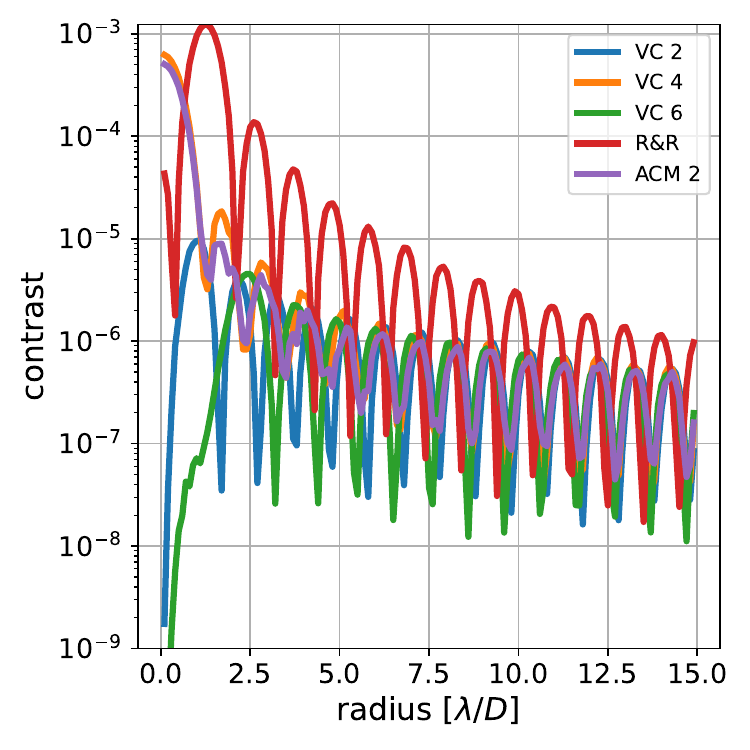}
    \caption{Azimuthally-averaged raw contrast in the SFP of the discretized coronagraphic system for five FPMs. The FPM features a spatial sampling of 10 pixels/($\lambda$/D) and a 64-bit grayscale phase resolution. The analysis assumes monochromatic light, an unobstructed entrance pupil, and no wavefront error.}
    \label{fig:singel_compared_10}
\end{figure}

To assess the contrast performance of various masks for a given sampling condition, Figure \ref{fig:singel_compared_10} compares the raw contrast performance of five different FPMs for a realistic scenario of 10 pixels/($\lambda$/D). This spatial sampling attempts to minimize aliasing without sacrificing too much field-of-view, assuming a finite number pixel budget. VC6 achieves the deepest contrast overall, closely followed by VC2, while R$\&$R generally delivers the worst contrast at nearly all angular separations. This is because R$\&$R achieves only partial on-axis starlight cancellation, as the $\pi$ phase-shifted core interferes destructively with the unshifted halo but does not completely remove the stellar light within the pupil \cite{roddier1997stellar}. All masks except R$\&$R reach similar contrast levels at angular separation beyond 2 $\lambda$/D or so, but VC4, VC8 (not shown on Figure \ref{fig:singel_compared_10} for clarity, due to similarity with VC4) and ACM2 exhibit the worst on-axis stellar rejection. FQPM has similar performance as VC2 (see Figure \ref{fig:res_spatial_sam}), thus it is not shown on Figure \ref{fig:singel_compared_10}. Because the considered FPMs have different off-axis transmission properties, the raw-contrast comparison should be interpreted together with the off-axis throughput curves shown in Appendix \ref{sec:off_axis_tp}. 

\subsection{Impact of limited FPM phase quantization (resolution)}
\label{sec:met_gs}
\begin{figure}[H]
    \centering
    \includegraphics[width=0.8\linewidth]{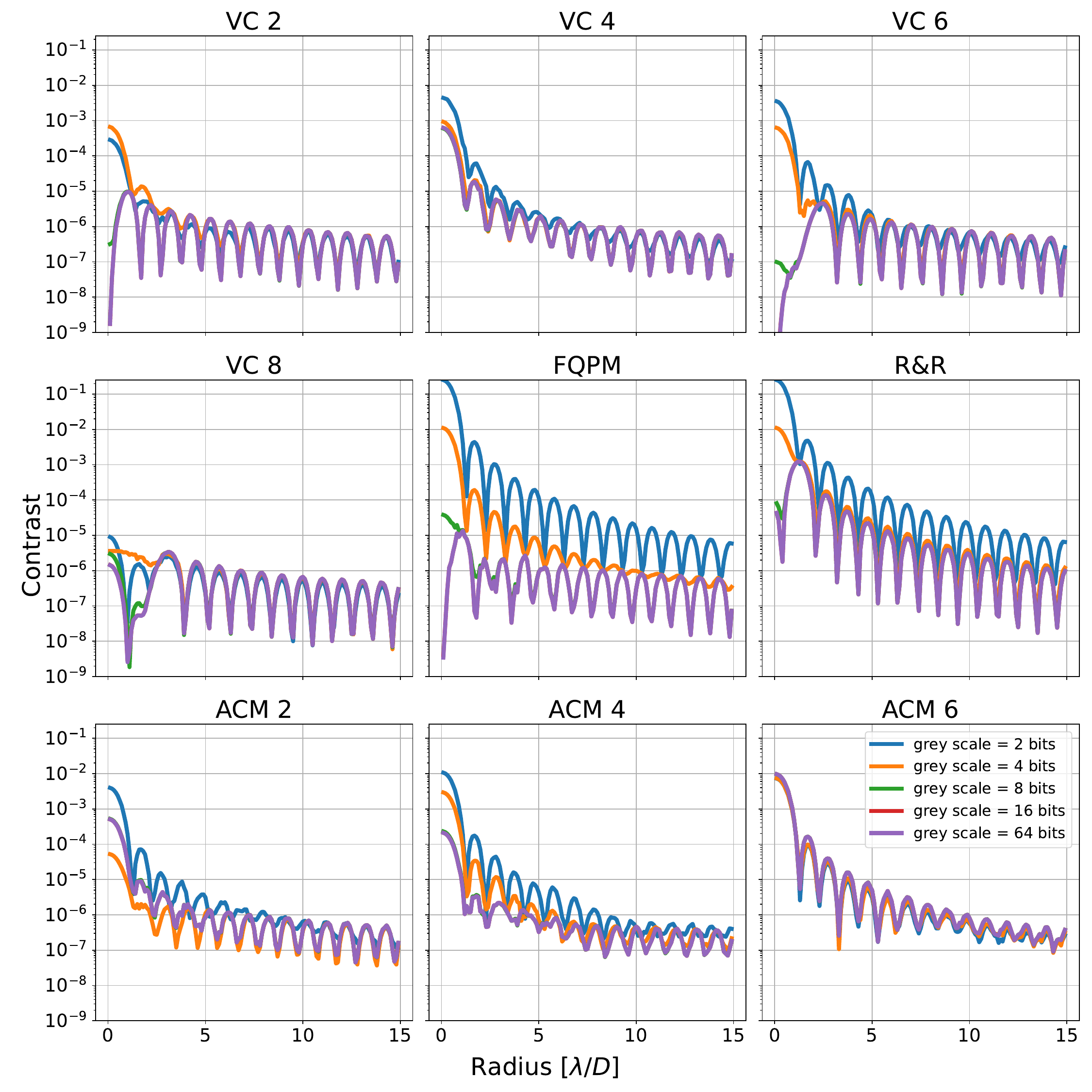}
    \caption{Azimuthally-averaged raw contrast in the SFP with various greyscale resolution levels, ranging from 2-bit to 64-bit (computational limit). Nine commonly used phase masks are evaluated. The entrance pupil is unobstructed. The FPM spatial sampling is set to 10 pixels/($\lambda$/D). Simulations assume monochromatic light and no wavefront errors.}
    \label{fig:greyscale}
\end{figure}

Another key parameter of discrete FPM coronagraphs is phase resolution or digitization. In this study, we assume a finite number of bits available to represent the full 2 $\pi$ phase range, from 2-bit (worst case scenario) to 64-bit (ideal case). As an example, most SLMs use a lookup table to linearly map 8- to 12-bit greyscale images to the full 2$\pi$ phase retardance range. Figure \ref{fig:greyscale} compares the raw contrast in SFP for various FPMs across multiple levels of quantization, ranging from 2- to 64-bit. The 64-bit represents the limit of our floating point precision. The 2-bit case, which represents only four discrete phase levels, results in significant degradation for nearly all FPMs, particularly at small angular separations where accurate phase modulation is critical. 4-bit resolution yields noticeable improvement but still falls short of theoretical contrast for most FPMs. 8-bit greyscale still yields slightly degraded performance, but only at the smallest angular separation, inside 2 to 3 $\lambda$/D. Higher-charge vortex masks (VC6 and VC8) are more sensitive to limited greyscale resolution. Designs relying on only a pair of phase values, such as FQPM and R$\&$R masks, are among the most affected FPMs when greyscale quantization is limited, but the performance drop tends to become negligible at and above 8-bits resolution. ACM masks in general are quite robust against limited greyscale quantization. Overall, an 8-bit resolution already provides close to ideal performance for the majority of FPM masks beyond 2-3 $\lambda$/D. We however note that those masks relying on only a handful of phase levels could theoretically retain their contrast performance, if the phase mapping was optimized to match those specific phase steps, and this even down to 2- to 4-bit quantization scenarios.

\subsection{Scenario with a centrally-obscured telescope aperture}
\label{sec:met_obs}
\begin{figure}[H]
    \centering
    \includegraphics[width=.8\linewidth]{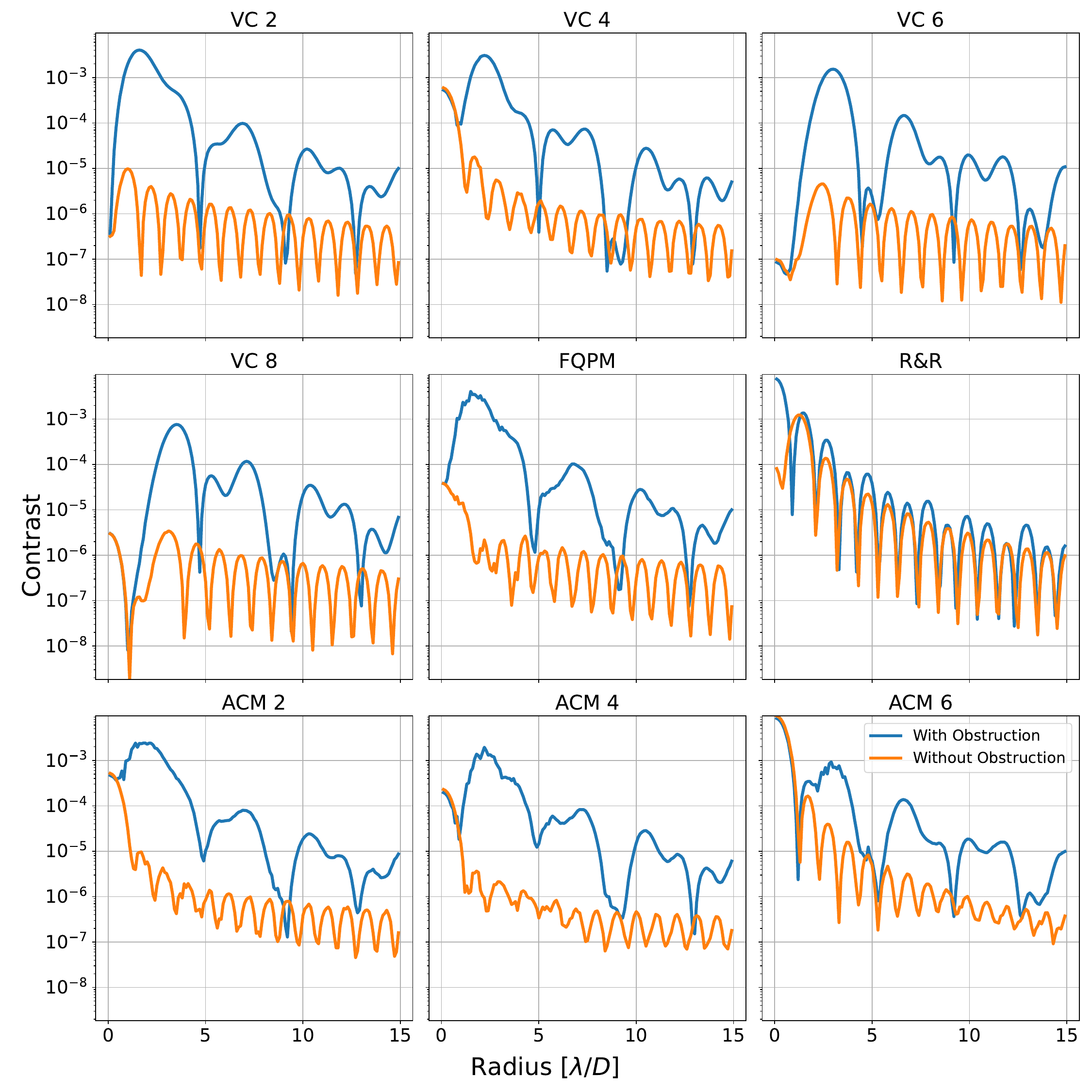}
    \caption{Azimuthally-averaged raw contrast in the SFP, comparing cases with and without a central obstruction in the entrance pupil. For the obstructed case, an LS secondary obstruction of $D'_s/D_s$ = 1 is used. The FPM has a spatial sampling of 10 pixels/($\lambda$/D) and an 8-bit grayscale phase resolution. Simulations assume monochromatic light and no wavefront errors.}
    \label{fig:obstruction}
\end{figure}

It is well known that the contrast and throughput performance of a coronagraphic system can be negatively affected by the presence of a central obstruction in the telescope entrance pupil, linked to the presence of a secondary mirror and its support structure. In order to study the interplay between this common contrast offender and the pixelated FPM parameters studied above, Figure \ref{fig:obstruction} shows contrast maps for various FPMs under centrally obstructed telescope pupil conditions, with the secondary mirror modeled as 25\% of the primary mirror diameter. The coronagraphic LS is configured as an exact replica of the entrance pupil, hence with neither over- nor under-sizing, to ensure a direct comparison among the various FPMs. Here, the FPM is sampled by 10 pixels/($\lambda$/D) with an 8-bit grayscale phase resolution, in order to illustrate a typical ``median scenario" that corresponds to a realistic SLM implementation (e.g. same parameters as in the DAG/PLACID instrument).

As shown in Figure \ref{fig:obstruction}, unless the central obstruction is explicitly taken into account in the coronagraph design \cite{oversize_LS_1}, its presence can significantly degrade the contrast performance by introducing additional diffraction and stellar leakage. The overall contrast level achieved by the R$\&$R consistently outperforms the other masks in presence of a central obstruction, especially at larger angular separations, though it admittedly had worse contrast performance than the other FPMs in the unobstructed case. The R$\&$R mask depends mainly on radial symmetry, and in the case with a central obscuration, this symmetry remains largely intact.

\subsection{Broadband contrast performance of discrete scalar FPMs}
\label{sec:met_bb}
\begin{figure}[H]
    \centering
    \includegraphics[width=1\linewidth]{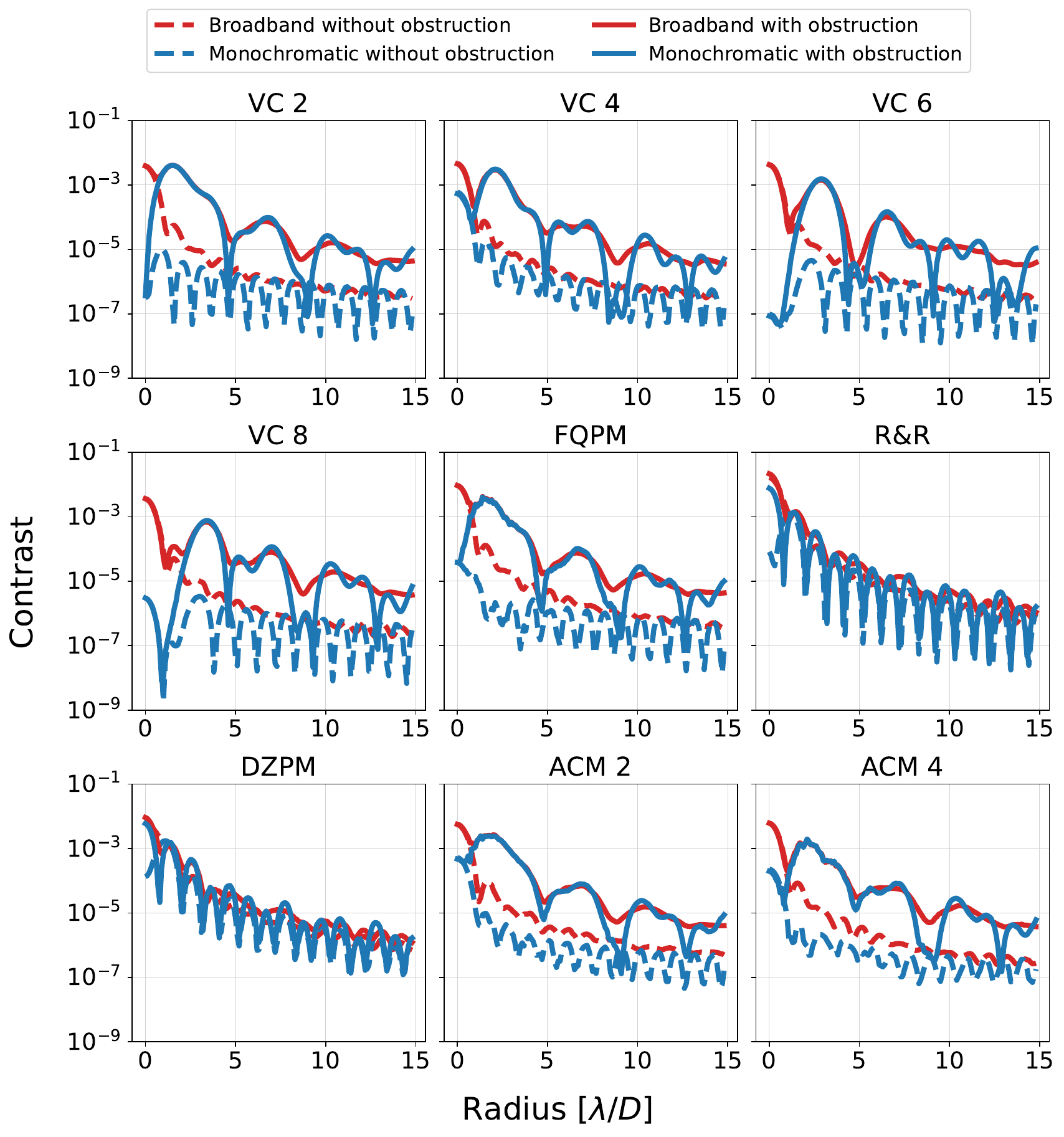}
    \caption{Azimuthally-averaged raw contrast in SFP under monochromatic or 20 $\%$ broadband conditions. Nine commonly used phase pixelated masks are evaluated. Dashed lines represent simulations with unobstructed entrance pupil, while solid lines correspond to cases including the 25 $\%$ entrance pupil obstruction ($D_s/D_p$ = 0.25) and an LS secondary obstruction of $D_s'/D_s$ = 1. The FPM has a spatial sampling of 10 pixels/($\lambda$/D) and an 8-bit grayscale phase resolution. Simulations assume no wavefront errors.}
    \label{fig:broadband_comparison}
\end{figure}

\begin{figure}[H]
    \centering
    \includegraphics[width=0.5\linewidth]{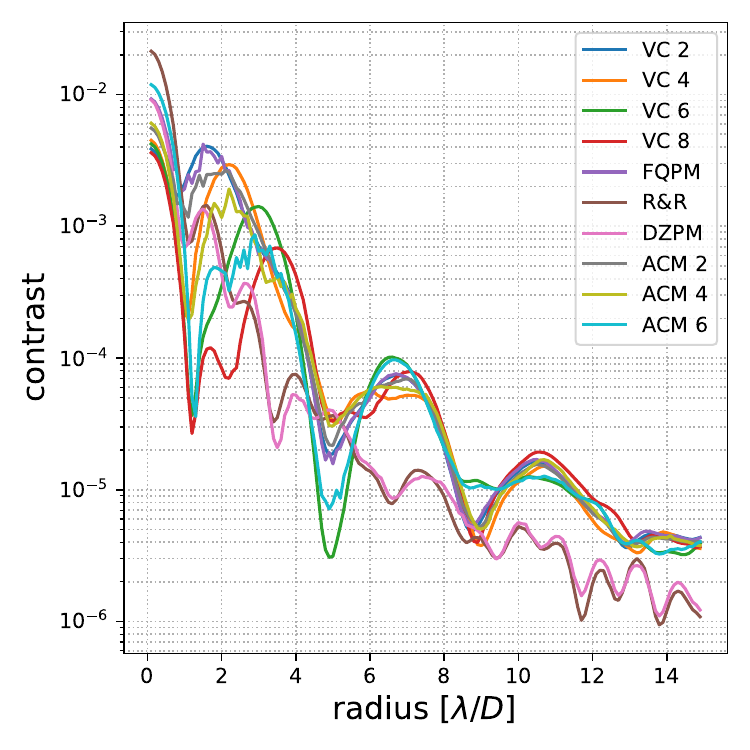}
    \caption{Azimuthally-averaged raw contrast in the SFP for various coronagraphic phase masks. The FPM features a spatial sampling of 10 pixels/($\lambda$/D) and an 8-bit grayscale phase resolution. Entrance pupil central obstruction of $D_s/D_p$ = 0.25 and Lyot stop secondary obstruction of $D_s'/D_s$ = 1. Simulations assume 20 $\%$ of broadband light and no wavefront errors.}
    \label{fig:broadband_combined}
\end{figure}

Typical astronomical window filters have a bandwidth of at least 20$\%$, but it is well known that most scalar FPMs can suffer significant loss in contrast performance when operating under broadband light. \cite{chromatic1,chromatic2} Figure \ref{fig:broadband_comparison} and \ref{fig:broadband_combined} plot the raw contrast curves for each FPM studied above for the default case of 10 pixels/($\lambda$/D) spatial sampling and 8-bit greyscale phase resolution (e.g. DAG/PLACID configuration with an SLM), with monochromatic or 20 $\%$ bandwidth broadband light. The cases with or without central obstruction are also considered. For the case without central obscuration, and given the already limited spatial and phase resolution of the FPMs, in general contrast performance only significantly deteriorates at small angular separation, typically inside 3 $\lambda$/D (FQPM exhibits slightly greater degradation within up to 5 $\lambda$/D under broadband conditions). Beyond these angular separations, broadband contrast tends to simply cancel out the PSF Airy rings, limiting the contrast loss as opposed to the monochromatic case. However, once a central obstruction is present in the EPP, the impact on contrast loss is major (up to 2 orders of magnitude) for most FPMs, and even up to very large angular separations (beyond 15 $\lambda$/D). The effect of broadband leakage becomes then comparatively limited and resides essentially within 2.5 $\lambda$/D or less; it is close to negligible at larger separations. In stark opposition, R$\&$R and DZPM patterns offer much greater overall contrast robustness, with respect to both broadband light conditions and presence of a central obstruction in the pupil. This is confirmed in Figure \ref{fig:broadband_combined}, where both FPMs clearly outperform their counterparts in the regime with both broadband light and central obstruction, at separation above 2 to 3 $\lambda$/D. Although the DZPM was originally developed as a more achromatic version of the classical R$\&$R mask, their performance remains nearly identical under the sampling condition of 10 pixels/($\lambda$/D), likely due to insufficient spatial resolution to fully resolve the dual zone phase structure (see Appendix Section \ref{sec:app_res_ene}). As discussed above, these raw-contrast trends do not by themselves provide a complete performance ranking, since the off-axis throughput differs between masks (see Appendix \ref{sec:off_axis_tp}).

\begin{figure}[H]
    \centering
    \includegraphics[width=0.9\linewidth]{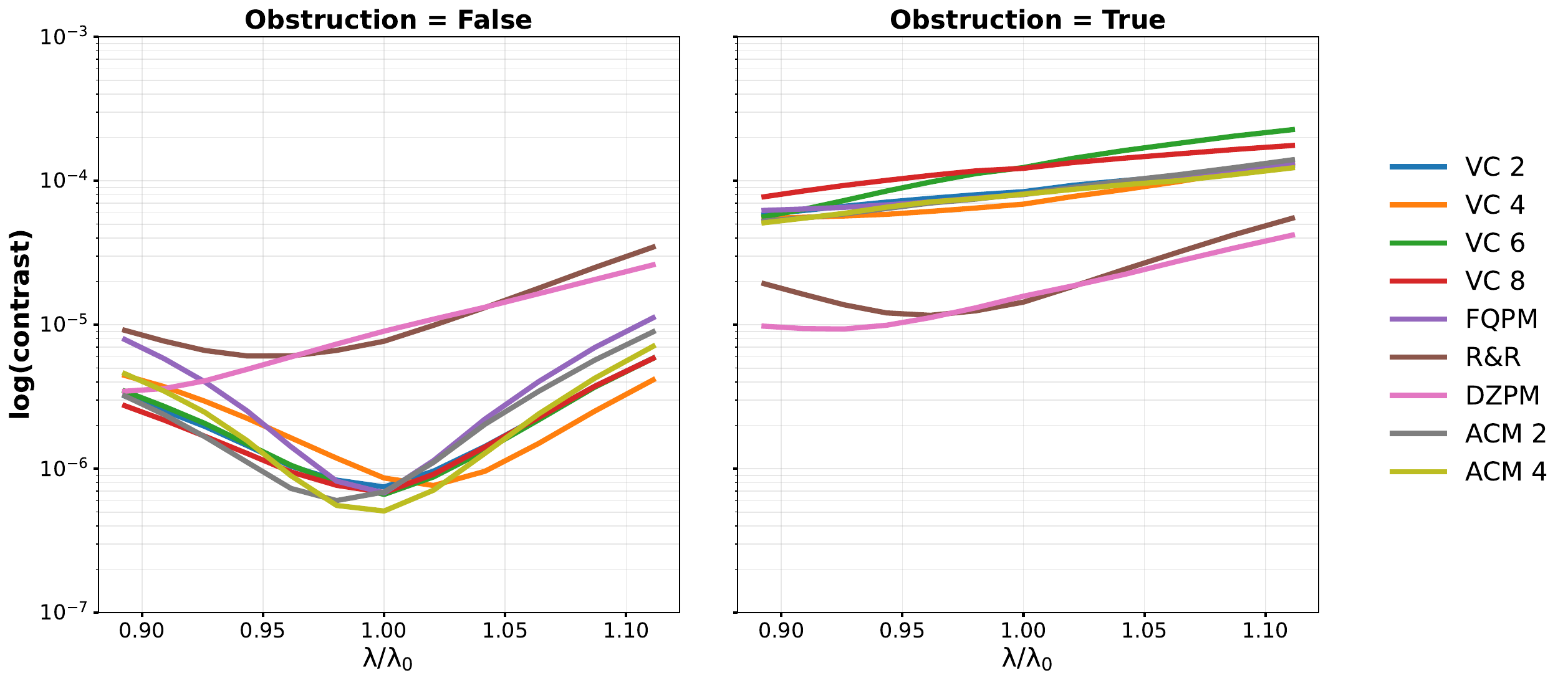}
    \caption{Mean raw azimuthally-averaged contrast in SFP calculated between 3 and 10 $\lambda/D$ as a function of wavelength, ranging from $0.9\,\lambda_0$ to $1.1\,\lambda_0$, with $\lambda_0$ the central wavelength.
    Nine commonly used phase masks are evaluated. The left panel shows results for an unobstructed entrance pupil, while the right panel shows results with central obstruction of $D_s/D_p$ = 0.25 and a Lyot stop secondary obstruction of $D_s'/D_s$ = 1. The FPM uses a spatial sampling of 10 pixels/($\lambda$/D) and an 8-bit grayscale phase resolution. Simulations assume no wavefront errors.}
    \label{fig:broadband_crosssection}
\end{figure}

To further evaluate the chromatic sensitivity of the discrete FPMs, the means of the azimuthally-averaged PSF raw contrasts are calculated between 3 and 10 $\lambda/D$ across a 20 $\%$ wavelength range from $0.9\,\lambda_0$ to $1.1\,\lambda_0$, with $\lambda_0$ being the central design wavelength. Here we consider discrete FPMs implemented with 10 pixels/($\lambda$/D) spatial sampling and 8-bit phase resolution again. As shown in Figure \ref{fig:broadband_crosssection}, the R$\&$R and DZPM masks are outperformed across the wavelength range, for the case without central obstruction. The slight redshift of the contrast minimum for VC4 is attributed to undersampling near the optical axis, as confirmed by simulations in ideal conditions, which is conducted at 100 pixels/($\lambda$/D) and 64-bit phase resolution (see Appendix Section \ref{app:mean_contrast}). However, when introducing a central obstruction, the R$\&$R and DZPM masks show a noticeable resilience to the presence of the obstruction. This evaluation step does not reveal a clear advantage of the DZPM over the R$\&$R design, at least when considering a spatial sampling of only 10 pixels/($\lambda$/D). However, when considering a more ideal case of 100 pixels/($\lambda$/D), the superior performance of the DZPM becomes evident, as shown in Appendix Section \ref{sec:app_res_ene} on Figure \ref{fig:resisual_ene}, which aligns with findings reported in the DZPM literature \cite{dual_zone}. In the interest of conciseness and to maintain relevance to a real-world implementation, notably the DAG/PLACID configuration operating with a H-band filter in its default configuration, we now restrict further analysis to the 20 $\%$ broadband light scenario. Unless otherwise noted, we also assume a spatial sampling of 10 pixels/($\lambda$/D) for the coronagraphic FPM and 8-bit phase quantization.

\subsection{Impact of the Lyot stop obstruction oversizing}
\label{sec:met_ls}
\begin{figure}[H]
    \centering
    \includegraphics[width=0.9\linewidth]{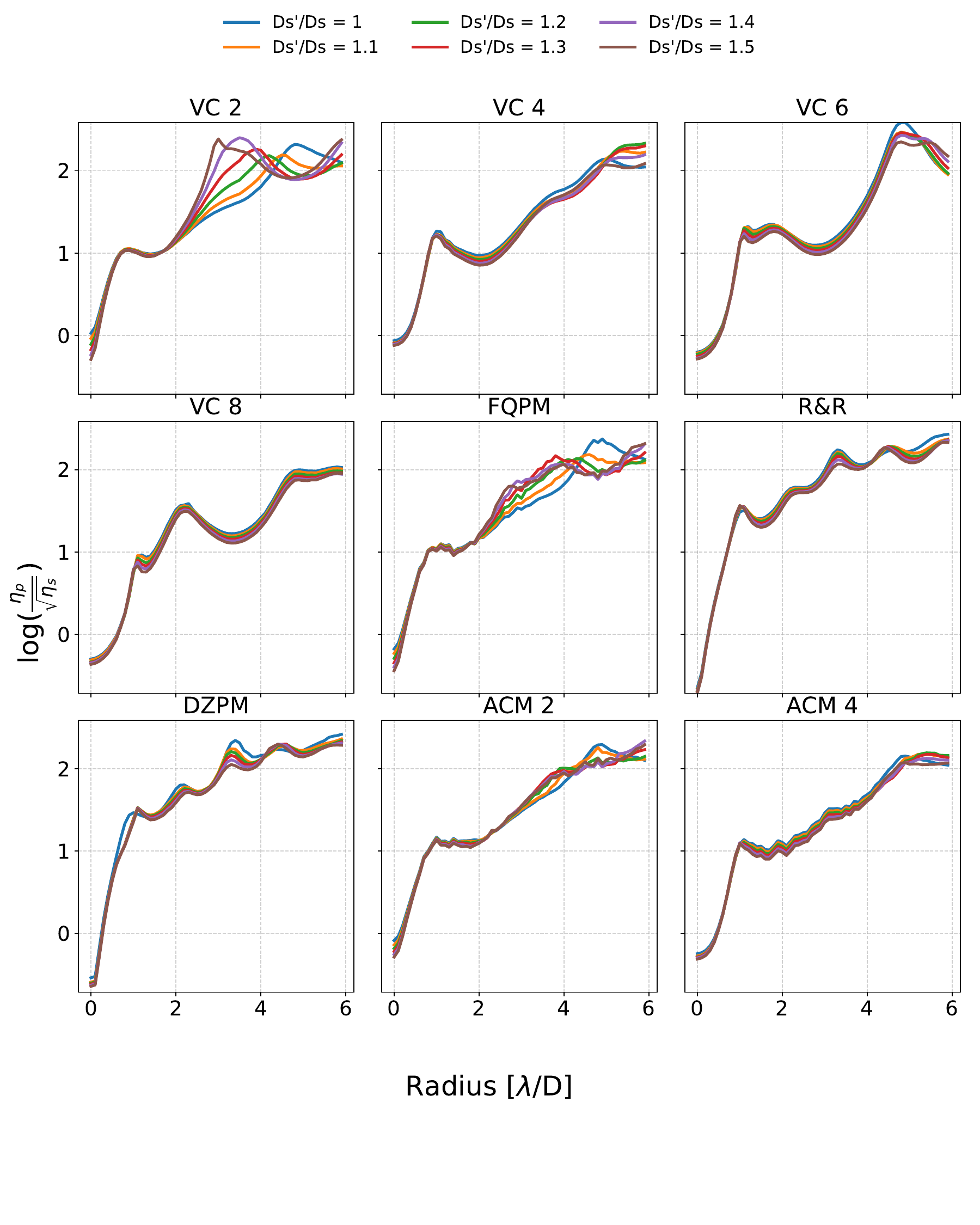}
    \caption{SNR ratio estimate $\epsilon$ = $\frac{\eta_p}{\sqrt{\eta_s}}$ for various coronagraph phase masks, evaluated under different LS secondary obstruction ratios ($D_s'/ D_s$). Nine commonly used phase masks are evaluated. Planet throughput is computed using horizontal shifts from the on-axis star, except for the FQPM, where a diagonal shift (45$^\circ$) is applied to avoid phase discontinuity. The entrance pupil obstruction size is set at 25 $\%$. The FPM features a spatial sampling of 10 pixels/($\lambda$/D) and an 8-bit grayscale phase resolution. Simulations assume 20 $\%$ of broadband light and no wavefront errors.}
    \label{fig:snr_lyot_stop}
\end{figure}

\begin{figure}[H]
    \centering
    \includegraphics[width=0.9\linewidth]{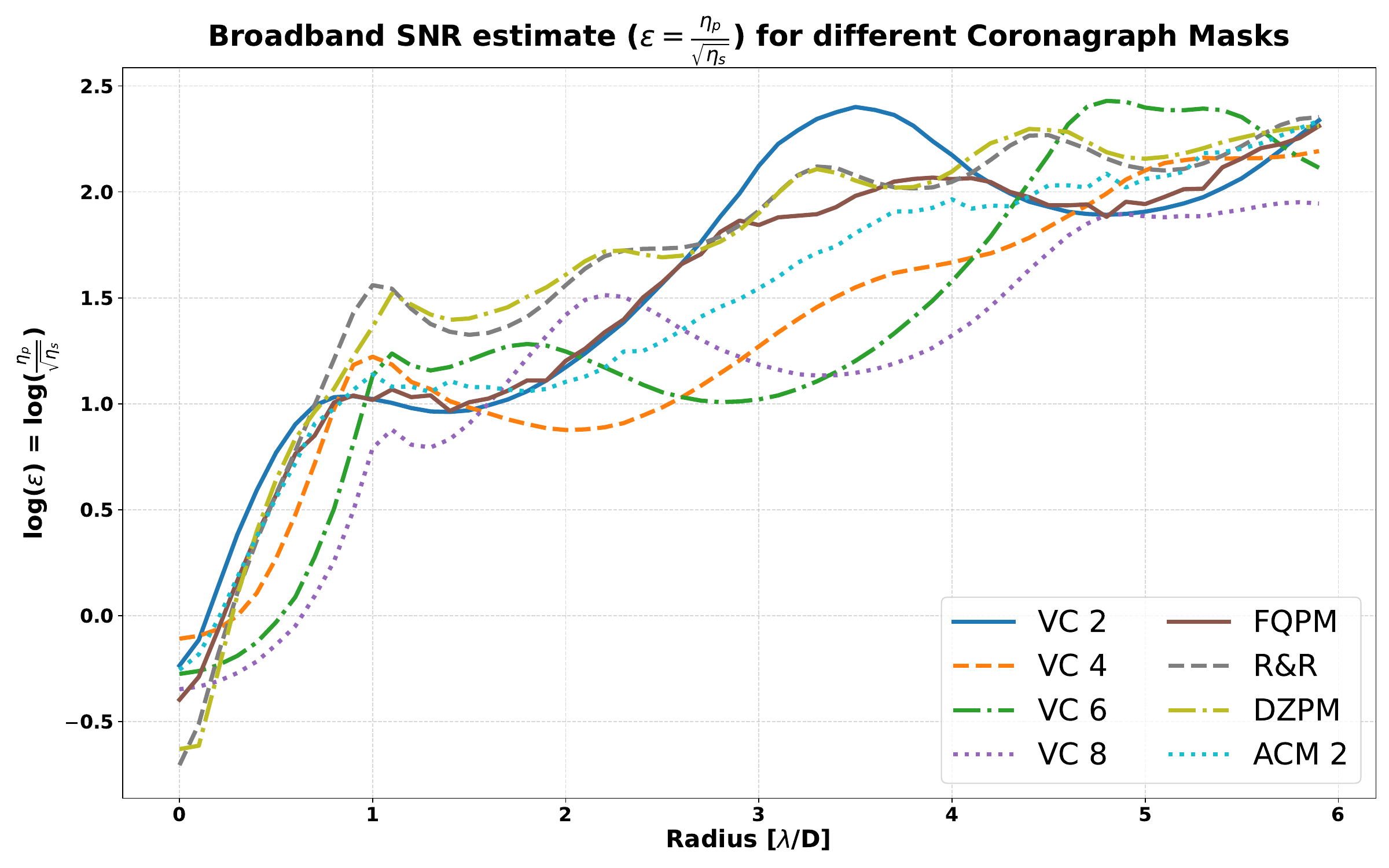}
    \caption{SNR estimates for various coronagraphs in realistic conditions 20$\%$ broadband light, DAG/PLACID
    telescope pupil with central obstruction: ($D_s/D_p$ = 0.25, $D_s'/ D_s$ = 1.4). The FPM features a spatial sampling of 10 pixels/($\lambda$/D) and an 8-bit grayscale phase resolution. Simulations assume no wavefront error.}
    \label{fig:snr_placid}
\end{figure}

We now consider a more realistic scenario, where the detrimental impact of the secondary mirror obstruction is addressed by an oversizing of the obscuration in the LS by up to 50 $\%$. As detailed in Section \ref{sec:tele_para}, the parameter $\epsilon$ = $\frac{\eta_{p}}{\sqrt{(\eta_{s})}}$ is used as a proxy to evaluate the corresponding impact on the SNR of a putative off-axis point source companion \cite{SNR_eqn}. As shown in Figure \ref{fig:snr_lyot_stop}, when oversizing the LS secondary obstruction ($D_s' / D_s > 1$), only low inner working angle (IWA) FPMs such as the VC2 and FQPM benefit from this scheme, with significant SNR gain in the 2 to 4 $\lambda$/D range, partially offset by a loss of SNR from 4 to 5 $\lambda$/D. In contrast, most of the other considered FPM designs remain relatively insensitive to LS secondary oversizing, within this specific 20 $\%$ broadband light scenario using a pixelated FPM with 10 pixels/($\lambda$/D) and 8-bit phase quantization. In cases where the SNR is independent of the LS central obstruction, the factor $D'_s/D_s$ should be mostly chosen based on the pupil stability budget of the telescope. Figure \ref{fig:snr_placid} presents the SNR estimate plot comparing the FPMs under realistic PLACID conditions with 20 $\%$ broadband light and $D_s'/D_s$ = 1.4. Between 0 and 2.5 $\lambda$/D, the R$\&$R and DZPM consistently achieve the highest SNR values, in other words, the best balance between contrast and throughput. From 3 to 4 $\lambda$/D, the VC2 mask performs the best among the designs with such an LS. While in the 4.5 to 6 $\lambda$/D range, VC6 mask shows the highest SNR. 

\subsection{Impact of pixel-level phase jitter, calibration errors and fill factor}\label{sec:met_pix_noise}
\begin{figure}[H]
    \centering
    \includegraphics[width=1\linewidth]{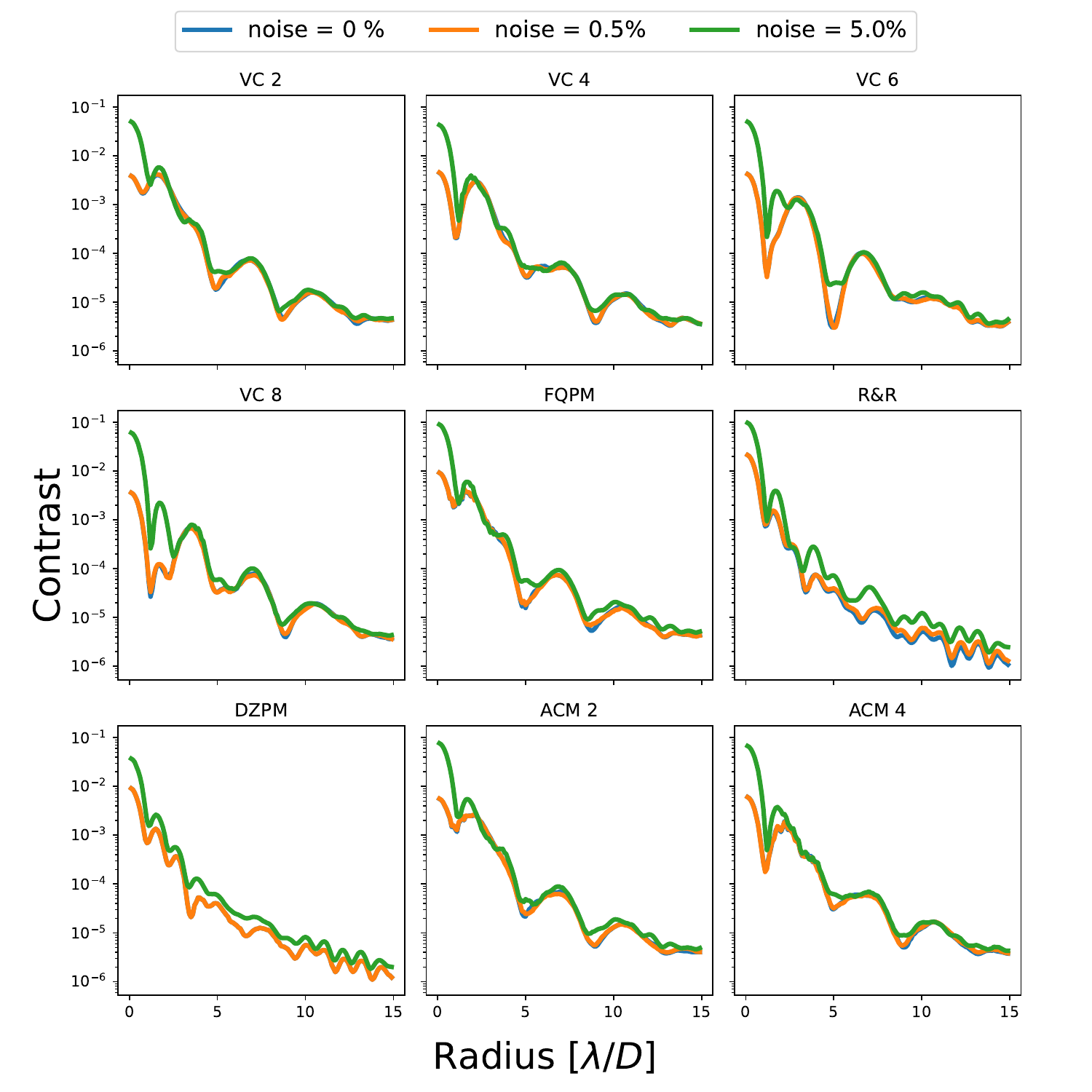}
    \caption{Azimuthally-averaged raw contrast in SFP with three different temporal pixel jitter levels: no noise, $\sigma$ = 0.5 $\%$ of 2 $\pi$, $\sigma$ = 5.0 $\%$ of 2 $\pi$. Nine commonly used phase masks are evaluated. The entrance pupil obstruction size $D_s/D_p$ is set at 25 $\%$. The FPM uses a spatial sampling of 10 pixels/($\lambda$/D) and an 8-bit grayscale phase resolution. Simulations assume 20 $\%$ of broadband light and no wavefront errors.}
    \label{fig:pixel_noise}
\end{figure}

Phase noise (spatial non-uniformity) at the single pixel level is introduced, under the assumption of a coronagraphic FPM sampled by 10 pixels/($\lambda$/D) with an 8-bit phase resolution. We also only focus on the realistic scenario with 20 $\%$ broadband light and a central obstruction of 25 $\%$. Based on the above results, we also introduce no oversizing of the secondary obscuration in the LS to avoid favoring an FPM versus another. As explained in Section \ref{sec:tem_phase_jitter}, each pixel value of the discrete FPM is assigned a phase offset error randomly drawn from a Gaussian noise distribution, and the same process is repeated 20 times with different random seeds to obtain statistically robust results. The standard deviation $\sigma$ of this phase offset distribution is expressed as a percentage of the maximum retardance of 2$\pi$ across all FPMs. Figure \ref{fig:pixel_noise} shows that the impact of pixel noise on the final PSF contrast is negligible, becoming significant only when the noise reaches about 5 $\%$ of 2$\pi$. This suggests that the phase masks considered here are generally robust to moderate pixel noise, at least for angular separations beyond 1 $\lambda$/D. The impact of the pixel noise is mostly concentrated within 1 $\lambda$/D, with vortex and ACM masks being most affected, particularly those with higher topological charges. This effect likely stems from their stronger reliance on increasingly fine phase sampling near the optical axis for optimal on-axis starlight suppression. In contrast, R$\&$R and DZPM maintain more stable performance, with less than an order of magnitude contrast loss on-axis. These can be explained by the fact that those FPMs rely only on 2 or 3 discrete phase values, hence the per-pixel phase noise effectively becomes spatially averaged. Overall, phase noise levels below 1$\%$, a typical value met by most SLM panels, result in no discernible contrast loss at any angular separation, at least in the considered cases with 10 pixels/($\lambda$/D) spatial sampling of the FPMs.

\begin{figure}[H]
    \centering
    \includegraphics[width=1\linewidth]{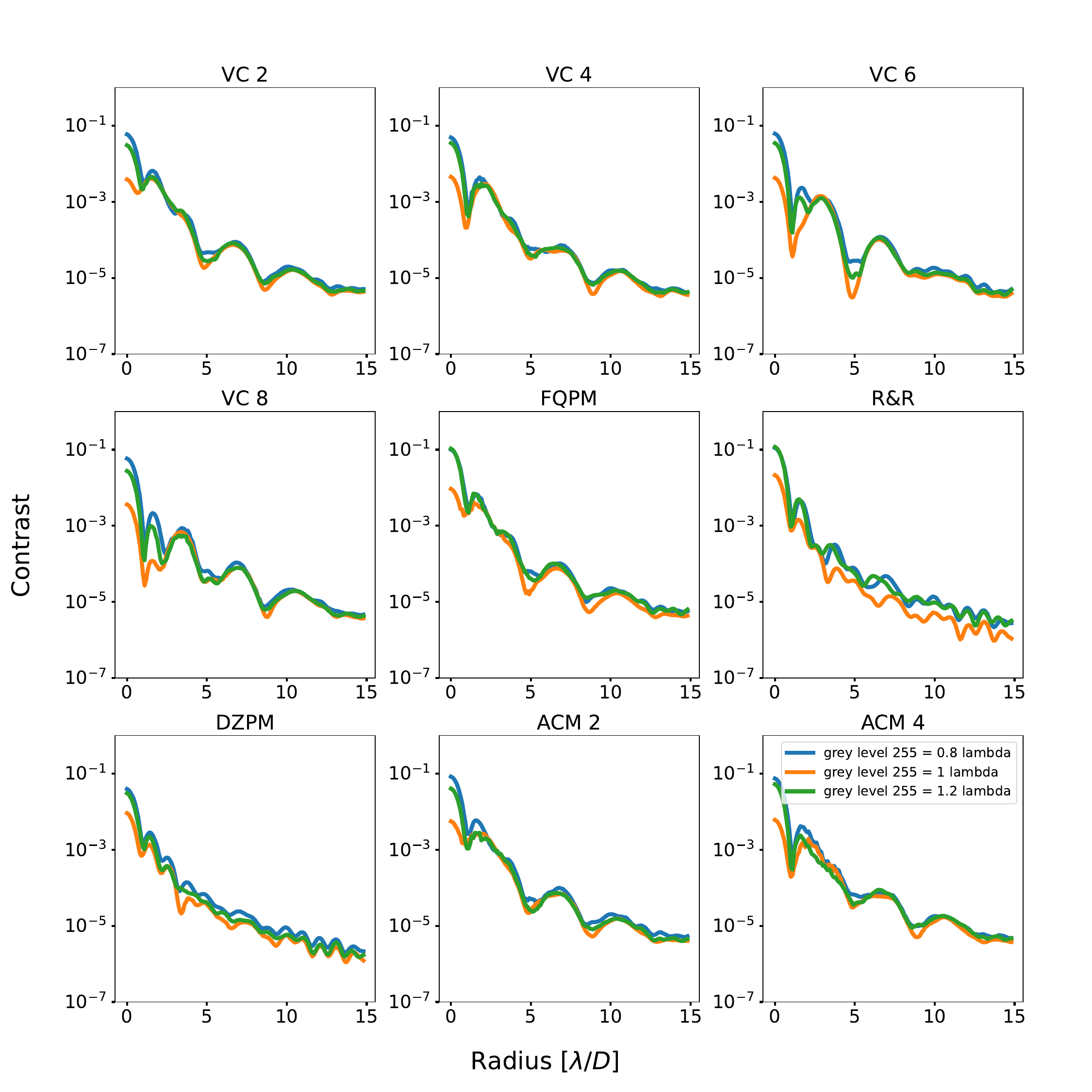}
    \caption{Azimuthally-averaged raw contrast in the SFP, simulated under phase calibration mismatch conditions in which a greyscale value of 256 corresponds to 0.8$\lambda_0$, 1$\lambda_0$ and 1.2$\lambda_0$. Nine commonly used phase masks are evaluated. The entrance pupil obstruction is set at 25 $\%$  and a Lyot stop secondary obstruction of $D_s'/D_s$ = 1. The FPM features a spatial sampling of 10 pixels/($\lambda$/D) and an 8-bit grayscale phase resolution. Simulations assume 20$\%$ broadband light and no wavefront errors.}
    \label{fig:cal_error}
\end{figure}

To evaluate how the accuracy of phase calibration in phase mask fabrication or implementation affects performance, we introduce slope errors into the phase maps by linearly modifying the grayscale-to-phase relationship. As shown in Figure \ref{fig:cal_error}, beyond 1 $\lambda$/D most masks demonstrate strong resilience to miscalibration, maintaining stable contrast profiles even with 20$\%$ deviation in calibration slope. 
However, subtle variations are nevertheless observed among the masks. VC6 and VC8 show greater sensitivity to phase calibration errors, as these high-charge masks rely on precise sampling of a rather fast phase slope for optimal performance. In contrast, ACM masks and DZPM perform robustly across all calibration scenarios. The DZPM slightly outperforms R$\&$R, likely due to its balanced achromatic phase mapping that provides broader tolerance to phase mismatch. Nevertheless, all FPMs experience a significant degradation in on-axis performance from mis-calibration errors, extending to 1 $\lambda$/D angular separation.

Finally, we also evaluate the impact of non-unity fill factor for the case of a pixelated implementation of an FPM, for example using an SLM. As introduced in Section \ref{sec:fill_factor}, our simulation tool enables up to 100 pixels/($\lambda$/D) sampling in the focal-plane. This allows us to artificially set the wavefront amplitude to zero along a row and a column of 10 focal-plane pixels, assuming the FPM is sampled by a discrete device (e.g. SLM) with a resolution of 10 pixels/($\lambda$/D). This scenario corresponds to about 81$\%$ fill factor, which is rather pessimistic, given that most currently available SLMs typically have fill factors well above 90$\%$. Still the impact of such a reduced fill factor on contrast performance is negligible regardless of angular separation, for all FPMs considered here. In order to ensure that we are not missing an effect at the second order level, and unlike the results shown above, the results in this simulation are generated for a more ideal scenario with monochromatic light and an unobstructed entrance pupil. The negligible role of the fill factor with respect to contrast is due to the fact that the corresponding light is diffracted outside the Lyot stop, where it is blocked, hence the impact is mostly limited to optical transmission (throughput).

\subsection{Combined impact of post-adaptive optics wavefront error residuals}
\label{sec:met_ao}

\begin{figure}[H]
    \centering
    \includegraphics[width=1\linewidth]{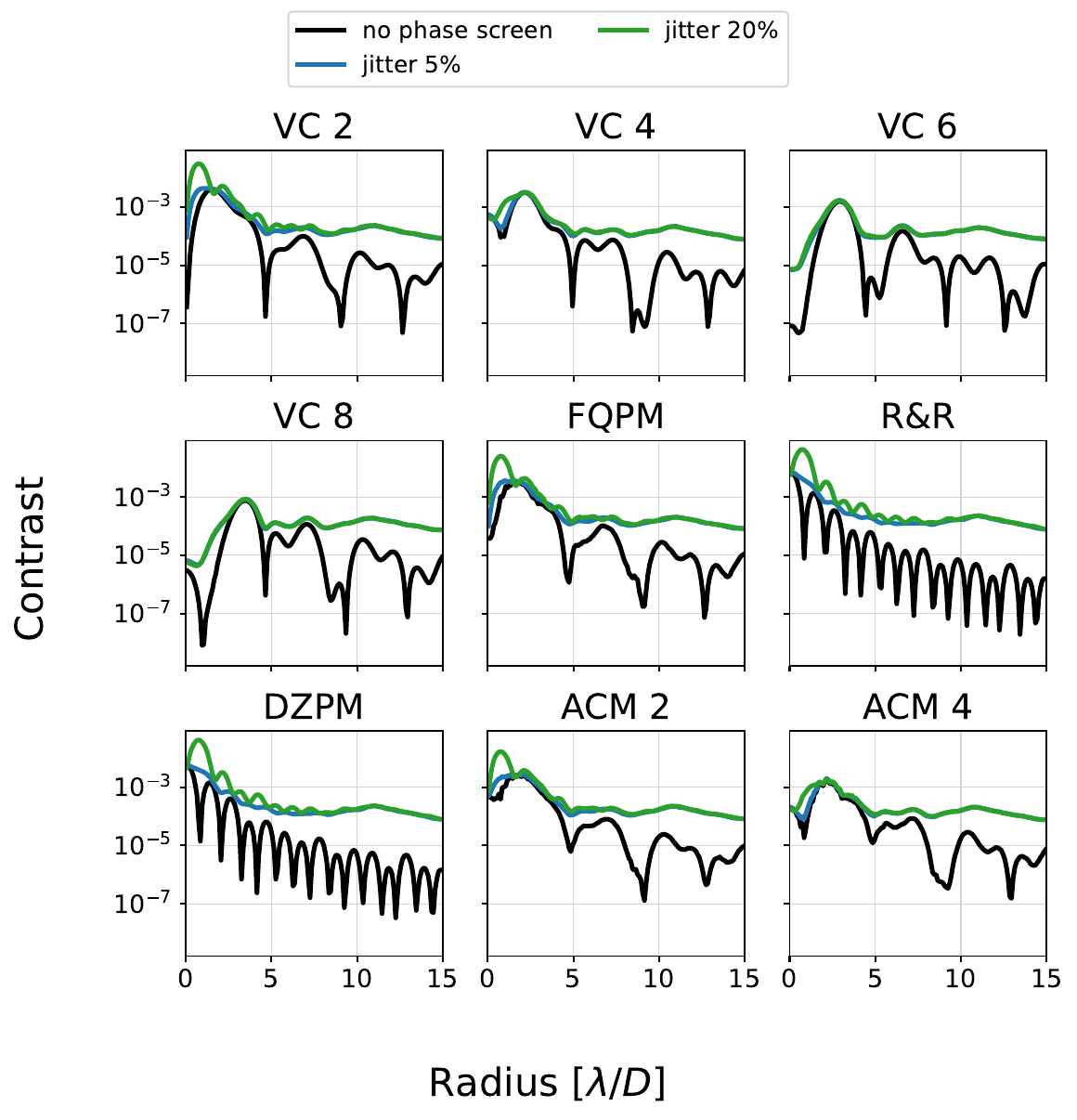}
    \caption{Azimuthally-averaged raw contrast in the SFP for nine commonly used phase masks. Comparing without post-AO residuals and with post-AO wavefront error residuals modeled using a phase screen series including additional low-order jitter levels of 5 $\%$ and 20 $\%$. $\lambda$/D rms The entrance pupil obstruction is set at 25 $\%$ and an LS secondary obstruction of $D_s'/D_s$ = 1. The FPM features a spatial sampling of 10 pixels/($\lambda$/D) and an 8-bit grayscale phase resolution. Simulations assume monochromatic light.}
    \label{fig:phase_screen}
\end{figure}

As detailed in Section~2.8, we subsequently introduce residual atmospheric wavefront errors from adaptive-optics simulations assuming a magnitude-8 star at a zenith angle of \(30^\circ\) under \(1''\) seeing conditions, together with residual tip/tilt jitter levels of 5\% and 20\% of $\lambda/D$ rms \cite{AO_jitter}, representative of idealized and realistic (e.g., VLT/SPHERE-like) scenarios, respectively. Figure~\ref{fig:phase_screen} shows that, relative to the case without AO residuals, the contrast performance of all FPMs is significantly degraded when these residuals are included. For vortex masks and ACMs, robustness to tip/tilt jitter increases with topological charge, at the expense of a larger inner working angle. The FQPM, R\&R, and DZPM patterns also undergo substantial contrast degradation in the presence of these AO residuals. The contrast bump around $11\,\lambda/D$ (see Figure \ref{fig:phase_screen}) is linked to the boundary of the influence region of the DAG/TROIA DM.

It should be emphasized that Figure \ref{fig:phase_screen} reports only the raw post-coronagraphic residual intensity, and does not account for any additional improvement (1 to 2 orders-of-magnitude contrast improvement) that may be achieved through post-processing techniques such as ADI, RDI, CDI, PCA-based analyses, or PSF-subtraction methods in general.

\section{Discussion}
This study attempts to provide a comprehensive overview of pixelated FPM coronagraphy under a range of realistic instrumental and observational parameters. Rather than relying on theoretical performance, the analysis focuses on how different pixelated coronagraphic FPMs are impacted by practical limitations such as finite spatial sampling, phase resolution, first-order calibration errors, but also more classical performance offenders such as chromaticity and a non-ideal telescope pupil (e.g. central obstruction). 

A key outcome of this work is the significant interplay between spatial sampling and the FPM type when evaluating achievable contrast performance. Masks with steep phase gradient or complex small-scale geometry, such as vortices with high charge, as well as R$\&$R and DZPM FPMs, require finer spatial sampling to approach their theoretical performance limits. At coarser sampling rates (e.g. $<$ 10 pixels/($\lambda$/D)), these masks exhibit increased starlight leakage, especially on-axis. Interestingly, this degradation is especially severe when the vortex charge is a multiple of four, as the four central pixels converge to an identical phase value (see Figure \ref{fig:dicrete_mask}), rather than producing an FQPM-like phase structure.

By contrast, the FQPM exhibits minimal sensitivity to spatial sampling in the coronagraphic focal plane, owing to its naturally square phase geometry that can easily accommodate reduced sampling. On the other hand, FQPM and R$\&$R are highly sensitive to phase quantization and calibration errors, because the masks rely on accurate and precisely defined half-wave ($\pi$) phase jumps. Thus even small deviations in phase encoding can significantly degrade the coronagraphic performance of FQPM and R$\&$R.

As expected from the scalar nature of the pixelated FPMs studied here, the impact of chromaticity on performance is severe across the board. The masks performing the best in the limited sampling and resolution conditions described above (vortices, FQPM and ACMs) tend to lose a fraction of their edge when compared to the others (R$\&$R and DZPM), but they still retain a clear advantage at small angular separations, especially for the higher topographic versions. When only raw contrast is considered, higher-charge vortex masks can appear favorable at small angular separations because of their stronger stellar suppression. However, this conclusion must be interpreted together with off-axis transmission. As shown by the SNR proxy in Figure \ref{fig:snr_placid}, once both stellar suppression and planet throughput are considered, no single mask is universally optimal, so the preferred choice depends on the separation and Lyot-stop configuration. However, the presence of a central obstruction in the telescope entrance pupil tends to change the state of play. Masks such as R$\&$R and DZPM prove to be quite robust in presence of a central obstruction, as the corresponding leakage is diffracted inside the secondary in the post-coronagraphic Lyot pupil plane (see Appendix Section \ref{sec:ap_lpp_im} Figure \ref{fig:lyot_plane}). This behavior holds even with broadband light (see Appendix Section \ref{app:mean_contrast} Figure \ref{fig:broadband_crosssection}), where both aforementioned FPMs clearly outperform their counterparts by up to an order of magnitude contrast from 3 to 10 $\lambda$/D. Adding a small oversizing of the secondary obstruction in the LS does not change the picture, as R$\&$R and DZPM keep their advantage in terms of SNR as well (see Figure \ref{fig:snr_lyot_stop}). It is notable that in this configuration only the VC2 mask truly benefits from an oversized secondary mask in the LPP, with the gain in contrast occuring primarily between 2 and 4  $\lambda$/D (see Figure \ref{fig:snr_lyot_stop}).


Another key aspect of the discrete implementation of coronagraphic FPMs is the precision and accuracy of the phase command at the single-pixel level. Here we observed that single-pixel phase precision (jitter) below 1$\%$ of one wave has no impact on achieved contrast for all FPMs considered here, presumably because the performance is already limited by the phase resolution of 8-bit in most of this study (0.39 $\%$ of 2$\pi$ error). Even larger phase jitter values of up to 5$\%$ result in moderate contrast degradation, mostly limited to an order of magnitude loss on-axis and within 1 $\lambda$/D. Similarly, phase accuracy (or calibration error) is also not a critical factor, when considering all the other limiting factors discussed here: 10 pixels/($\lambda$/D) spatial sampling, 8-bit resolution, 20$\%$ broadband light, 25$\%$ central obstruction. Even with up to 20$\%$ phase slope miscalibration, the resulting contrast loss remains essentially concentrated within 1 $\lambda$/D for all FPMs. An exception occurs for higher topographic charge vortices, where the impact extends to about 2.5 $\lambda$/D (see Figure \ref{fig:cal_error}). We also confirmed that fill factor limitations essentially result in reduced throughput, with negligible impact on contrast.

Finally, to better understand the interplay between the actual contrast performance of pixelated scalar phase masks and the level of wavefront-error residuals, we inject a model of post-adaptive-optics wavefront-error residuals for the TROIA AO system of the DAG telescope\cite{AO_simulation, AO_simulation_2}, together with residual tip/tilt jitter terms (jitter levels of 5\% or 20\% of $\lambda/D$ rms) \cite{AO_jitter}. As expected, and as illustrated in Figure \ref{fig:phase_screen}, these effects introduce leakage terms that quickly dominate the raw-contrast budget compared with the sampling effects and chromatic limitations discussed above. However, this does not imply that the choice of FPM becomes irrelevant. On the contrary, as shown in Figure \ref{fig:phase_screen}, different FPM designs do not respond identically to these non-ideal conditions: vortex masks and ACMs become increasingly sensitive to tip/tilt jitter as the topological charge decreases, while FQPM, R\&R, and DZPM also experience substantial degradation, as is the case for all masks providing a small inner working angle. Thus, although adaptive-optics residuals set the raw-contrast floor in a realistic ground-based telescope environment, the choice of focal-plane mask still governs the coronagraph robustness to residual low-order aberrations and jitter, as well as the inner working angle and the throughput for off-axis sources.

An important implication of these results is that the performance of pixelated discrete FPMs does not appear to be primarily limited by spatial sampling, phase quantization, or moderate pixel-level noise. Instead, the dominant limitation arises from the non-ideal telescope aperture (Figure \ref{fig:obstruction} and \ref{fig:broadband_comparison}), and to a lesser extent from the scalar nature of the imprinted phase modulation (Figure \ref{fig:broadband_comparison}), which leads to significant chromatic leakage under broadband illumination. This suggests that future progress in adaptive coronagraphy may depend less on incremental improvements in pixel resolution and phase addressing, but rather more on the development of active phase modulation panel technologies with achromatic behavior. In that respect, any developments towards geometric-phase SLMs\cite{geometric_SLM_1, geometric_SLM_2} or reduced-pitch DMs, could represent a future avenue for broadband active focal-plane coronagraphy. In addition, complementary system-level approaches may help mitigate chromatic leakage, including achromatic scalar phase-mask concepts that exploit a larger phase retardance, as in the wrapped-vortex approach \cite{wrap_vortex}.

\section{Conclusions}

In this study, we conducted a comprehensive theoretical analysis of active FPM coronagraphy using discrete, pixelated devices. Although the present analysis is motivated by the fact that LCoS SLMs are used in the DAG/PLACID instrument, its implications extend more broadly to pixelated programmable phase-mask technologies. In particular, the results may inform the development of future devices with higher spatial resolution, as well as alternative architectures such as deformable mirrors with smaller actuator pitch. We systematically evaluated the impact of key device parameters such as spatial sampling, phase quantization, calibration precision and accuracy, as well as fill factor. 

Our findings highlight that the choice of coronagraphic FPM significantly influences sensitivity to discretization effects, at least in ideal conditions (no wavefront errors).  Vortex and ACM, particularly at higher topological charges, require higher spatial resolution (at least 10 pixels/($\lambda$/D) and phase resolution (8-bit or above) to maintain near-ideal contrast performance, especially at small working angles, while FPMs relying on only 2 or 3 phase levels (such as FQPM, R$\&$R and DZPM) are more robust in this regime.  In contrast, when more realistic instrumental conditions are considered, such as broadband light (20 $\%$ bandwidth) and a centrally obstructed entrance pupil, radial masks like R$\&$R and DZPM demonstrate robust and resilient contrast performance. They also offer a clear advantage in achievable SNR due to their high transmission, given that they do not require oversizing of the secondary mask in the Lyot stop. 

When considering a ground-based implementation, most of these results are ultimately overshadowed by the presence of post-adaptive optics wavefront error residuals, which clearly dominate the contrast and SNR budget by one to two orders of magnitude. Hence a practical implementation of an active programmable FPM coronagraph from the ground may be mostly driven by real-time considerations related to observing conditions or science requirements. We can mention two examples here. First, an observer may decide to use a higher charge FPM (vortex or ACM) in the case of detrimental weather conditions, that cause large residual tip/tilt jitter. This choice allows observation to proceed at all, though at the acceptable cost of losing some inner-working angle discovery space. A second example, driven by science considerations, would see an observer retain a R$\&$R or DZPM coronagraph to straightforwardly implement a ``dual-star" coronagraph, enabling the observation of binary (or higher order multiples) stars; or pick a high charge FPM to be able to observe a giant spatially resolved star. In addition, when considering spatial sampling in a real-world implementation with a finite number of pixels (e.g. a few Mpixels for most SLMs), one usually has to trade usable on-sky field-of-view versus the number of pixels per telescope beam-width, making this choice a key parameter when considering the overall system design.

 Overall, our simulations show that using commercially available LCoS SLMs with 8-bit phase sampling and about 10 SLM pixels per telescope diffraction beam is a viable scheme for implementing a programmable active focal-plane coronagraph on a ground-based facility. Indeed, the presence of the secondary mirror central obstruction and post-adaptive wavefront error residuals overwhelmingly dominate the contrast budget, even considering the scalar nature of the phase delay introduced by LCoS SLMs. Currently, the main trade-off is between the poor transmission (strictly less than 50 $\%$) of an SLM-based instrument requiring linearly-polarized light, and the extra freedom offered by the programmable nature of such an active coronagraph. We intend to evaluate this approach on-sky with the PLACID instrument for the 4-m DAG telescope, currently in the Assembly, Integration and Validation (AIV) phase on the Nasmyth platform, with first light foreseen later in 2026.

In parallel, we are currently running an extensive interferometric characterization campaign of SLMs using digital holographic microscopy (DHM), to refine the models introduced in this work with more realistic parameters derived from real SLM measurements. This effort is not limited to LCoS SLMs readily available on the market, but will also address novel and promising SLM or DM technologies currently being introduced, such as opto-thermal SLM (T-SLM) \cite{PhotothermalSLM} approaches that may be used to generate an achromatic phase shift for unpolarized light. This effort is important not only for the future evolution of programmable focal-plane active coronagraphy, for instance on space-born observatories, but also for the next-generation of wavefront control devices that will play a critical role in achieving the sub-nanometer wavefront control accuracy and precision required to image exo-Earths around solar-type stars, either from the ground (e.g. ELT/PCS) or in space (e.g. HWO).

\newpage
\appendix

\section{Off-axis throughput of the considered focal-plane phase masks}
\label{sec:off_axis_tp}

\begin{figure}[H]
        \centering
        \includegraphics[width=0.8\linewidth]{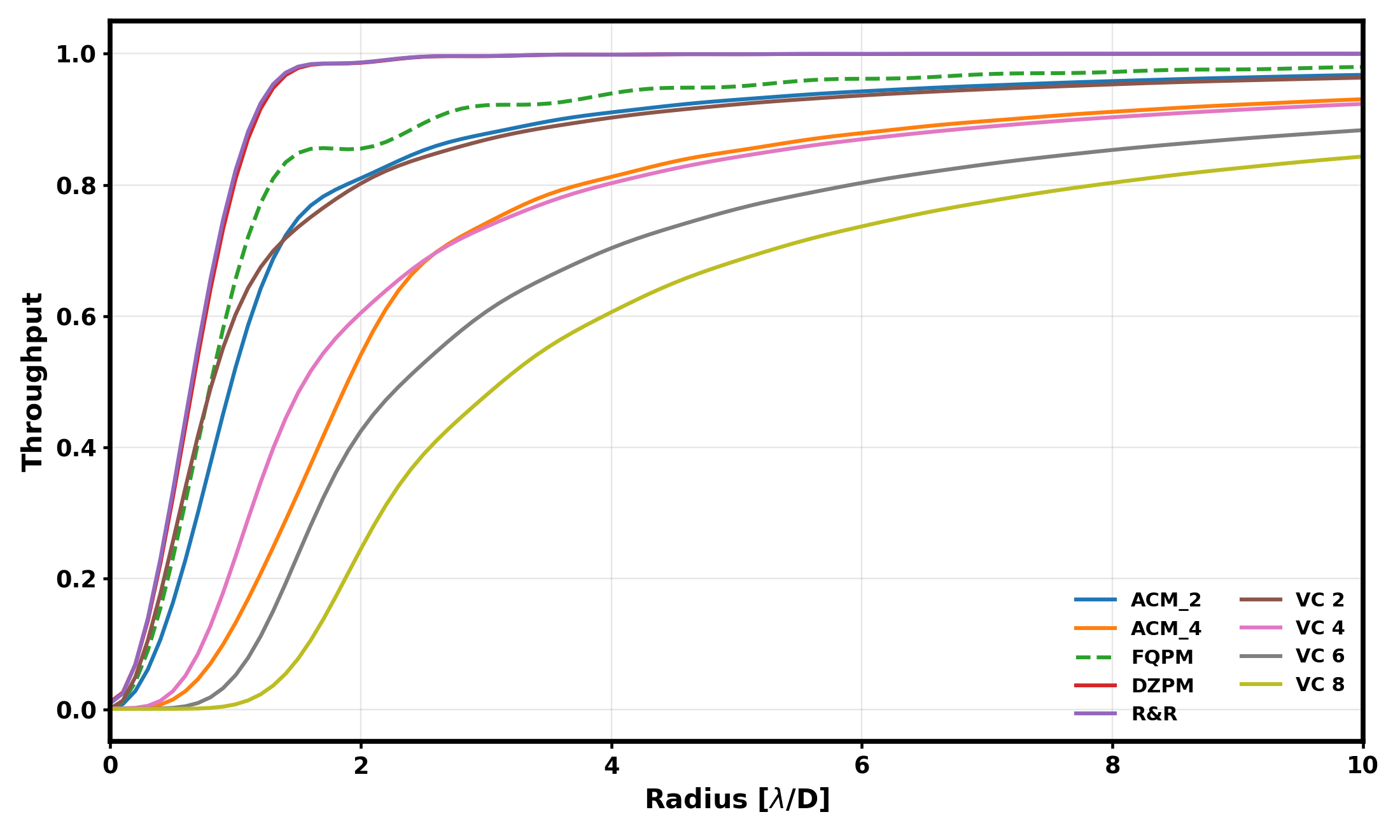}
        \caption{Off-axis throughput as a function of angular separation for the FPMs considered in this work. The curves are computed for a representative configuration with 10 pixels per $\lambda/D$ FPM sampling, 8-bit phase quantization, monochromatic light, and an unobstructed entrance pupil. The throughput is normalized to the off-axis PSF obtained without a coronagraph. These throughput curves provide complementary information to the raw-contrast comparisons in Figures ~\ref{fig:broadband_combined} and \ref{fig:broadband_crosssection}.}
        \label{fig:tp}
    \end{figure}

Figure \ref{fig:tp} shows the off-axis throughput as a function of angular separation for all FPMs considered in this work, computed for the same representative configuration used in the main comparison: 10 pixels per ($\lambda/D$) spatial sampling, 8-bit phase quantization, monochromatic light, and an unobstructed entrance pupil. The throughput is normalized to the off-axis PSF energy obtained without a coronagraph.

The different masks exhibit significantly different inner-throughput behavior. The R \& R and DZPM masks reach high throughput close to (1~$\lambda/D$), while the FQPM and low-order ACM/vortex masks show intermediate behavior. Higher-charge vortex masks, especially VC6 and VC8, have substantially lower throughput at small angular separations and recover more gradually with separation. This confirms that raw contrast curves alone should not be used to rank the coronagraphs, particularly at small separations, since masks providing stronger stellar suppression can also reduce the transmitted planet signal. The throughput curves therefore provide complementary information for interpreting the contrast-only comparisons in Figure \ref{fig:broadband_combined} and \ref{fig:broadband_comparison}.

\section{Illustration of low phase resolution}
\label{sec:ap_gs_2bit}

\begin{figure}[H]
\begin{center}
\begin{tabular}{c}
\includegraphics[width=0.8\linewidth]{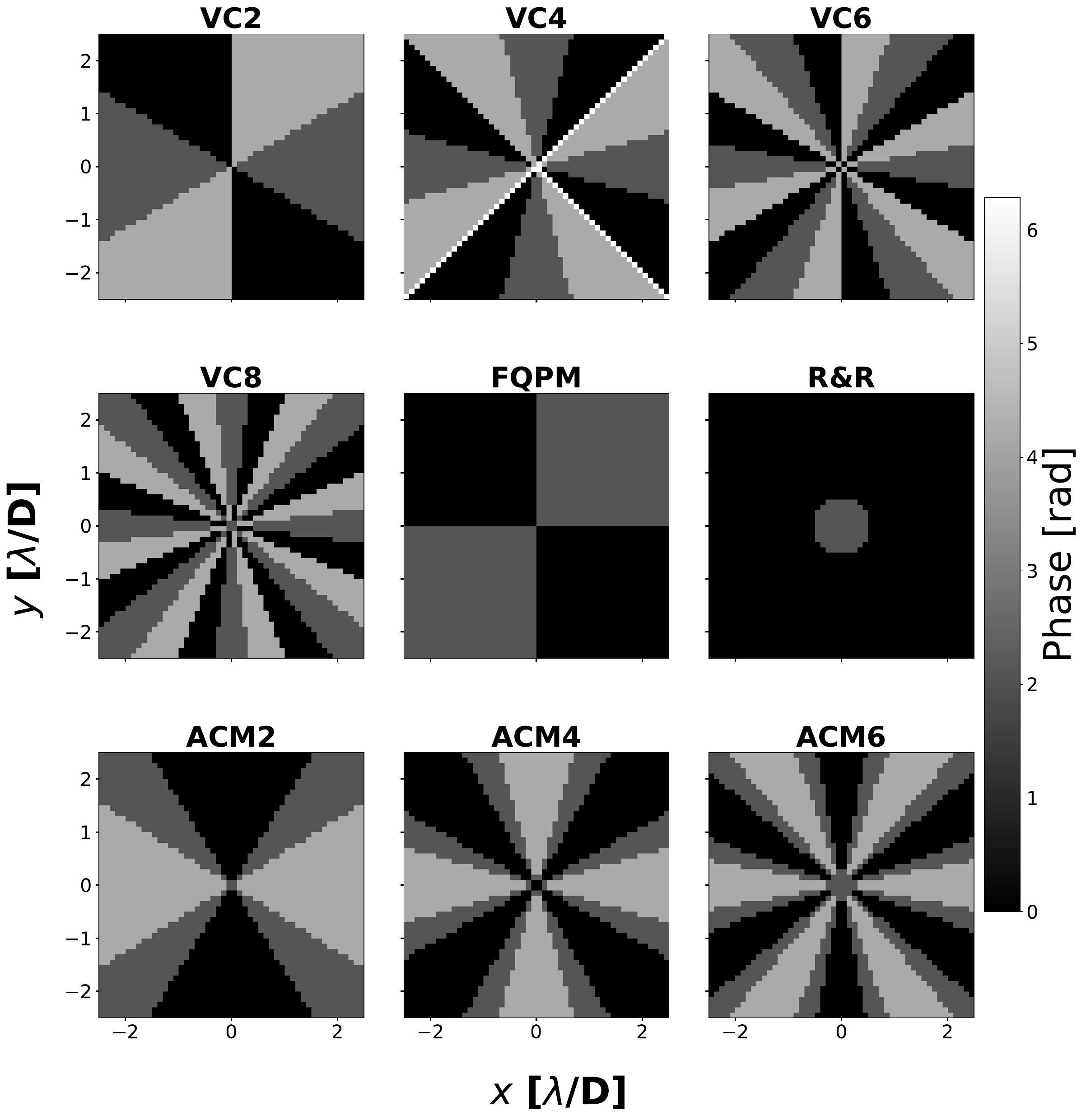}\end{tabular}
\end{center}
\caption 
{ \label{fig:greyscale_2bit}
Illustration of discrete FPMs with a spatial sampling of 10 pixel/($\lambda$/D) and phase resolution of 2-bit.} 
\end{figure}

To illustrate why even very low phase resolution can still provide decent coronagraphic performance, Figure \ref{fig:greyscale_2bit} shows  several phase masks with extremely low phase quantization represented by  the 2-bit case. Even at 2-bit resolution, the quantized masks preserve the overall azimuthal phase ramp and central phase discontinuity responsible for destructive interference at small angular separations. For a spatial sampling of 10~pixels$/(\lambda/D)$, the geometric representation of the central singularity remains unchanged across all bit depths. What changes as the phase resolution decreases is the stepwise approximation of the azimuthal ramp. These discrete phase steps introduce additional phase errors, which increase the residual stellar leakage and therefore limit the achievable null depth. Nevertheless, because the overall symmetry is preserved, significant suppression remains possible at moderate angular separations even for extremely low phase resolution implementations.

The fact that most of the 2-bit masks exhibit only three phase values, rather than four, arises from the phase quantization convention adopted in the simulation. In the simulation, we assume that $2\pi$ can be represented exactly, so that a 2-bit mask is described by the fourth phase levels $[0,\;2\pi/3,\;4\pi/3,\;2\pi]$. This is notably to match the real-life scenario of using an SLM panel, where the entire grey level range is mapped to one wave of retardance. Therefore phase values in the interval $[0,\;2\pi/3)$ are mapped to $0$, values in $[2\pi/3,\;4\pi/3)$ are mapped to $2\pi/3$, and so forth, with only values exactly equal to $2\pi$ mapped to $2\pi$. In practice, because of the finite spatial sampling, many of the discretized vortex masks do not actually reach a phase value of exactly $2\pi$, so most of the corresponding 2-bit patterns contain only three distinct phase values. This effect is most visible in the extreme low-resolution case of 2-bit quantization and becomes negligible for more realistic cases such as 8-bit phase resolution.

\section{Effect of sub-pixel mask centering on coronagraphic performance}
\label{sec:ap_sub_pixel}
\begin{figure}[H]
\begin{center}
\begin{tabular}{c}
\includegraphics[width=0.8\linewidth]{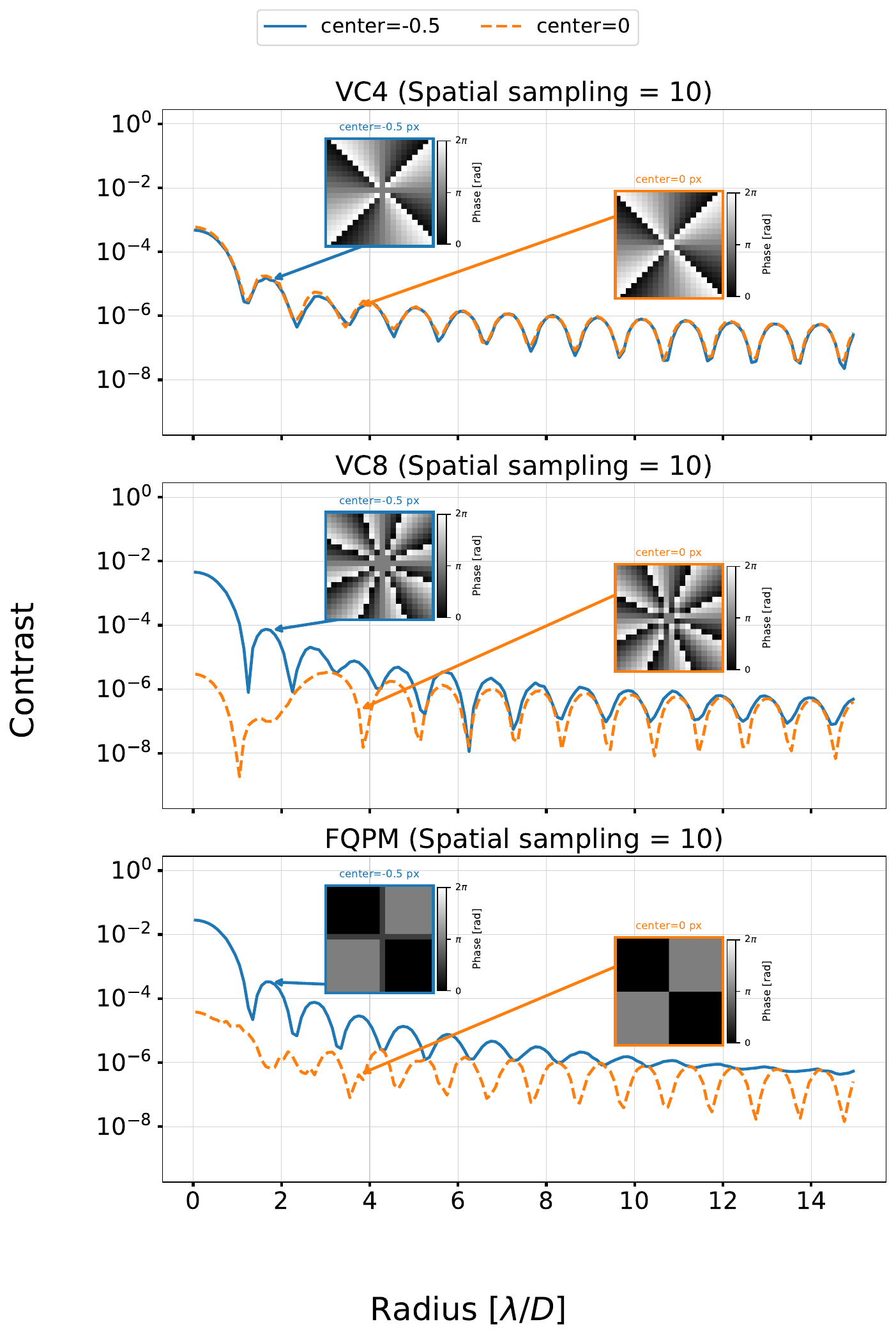}\end{tabular}
\end{center}
\caption 
{ \label{fig:mask_centering}
Azimuthally-averaged raw contrast curves in the SFP for discretized coronagraphic systems. For FQPM, VC4 and VC8, configurations with centers shifted by a fractional pixel (blue) are compared with the original centered configurations (yellow). The analysis assumes monochromatic light, an unobstructed entrance pupil, 10 pixels/($\lambda$/D) spatial sampling and 64-bit greyscale phase resolution
for the FPMs and no wavefront error.} 
\end{figure}
 The centering of the mask on the discrete pixel grid can significantly affect the coronagraphic performance, although the impact depends strongly on the mask design. To illustrate this effect, the centers of the FQPM, VC4, and VC8 masks, as well as the PSF core, were shifted by half a pixel, and their contrast performance was compared with that of the original configurations. By ``original'' we mean the default case in which the center of the grid lies at the intersection of the four central pixels. As can be seen on Figure \ref{fig:mask_centering}, for VC4, the sub-pixel shift breaks the original four-zero-pixel pattern at the singularity and replaces it with a cross-like central structure. However, this modified pattern leads only to a negligible improvement in contrast. In the case of VC8, the fractional-pixel shift creates additional zero-valued pixels around the singularity, which considerably worsen the contrast performance. The effect is much more pronounced for the FQPM: a sub-pixel shift effectively introduces a central cross-like feature with intermediate phase values around $\pi/2$, causing a substantial contrast degradation compared with the nominal configuration.

\section{LPP Image}
\label{sec:ap_lpp_im}
\begin{figure}[H]
\begin{center}
\begin{tabular}{c}
\includegraphics[width=1\linewidth]{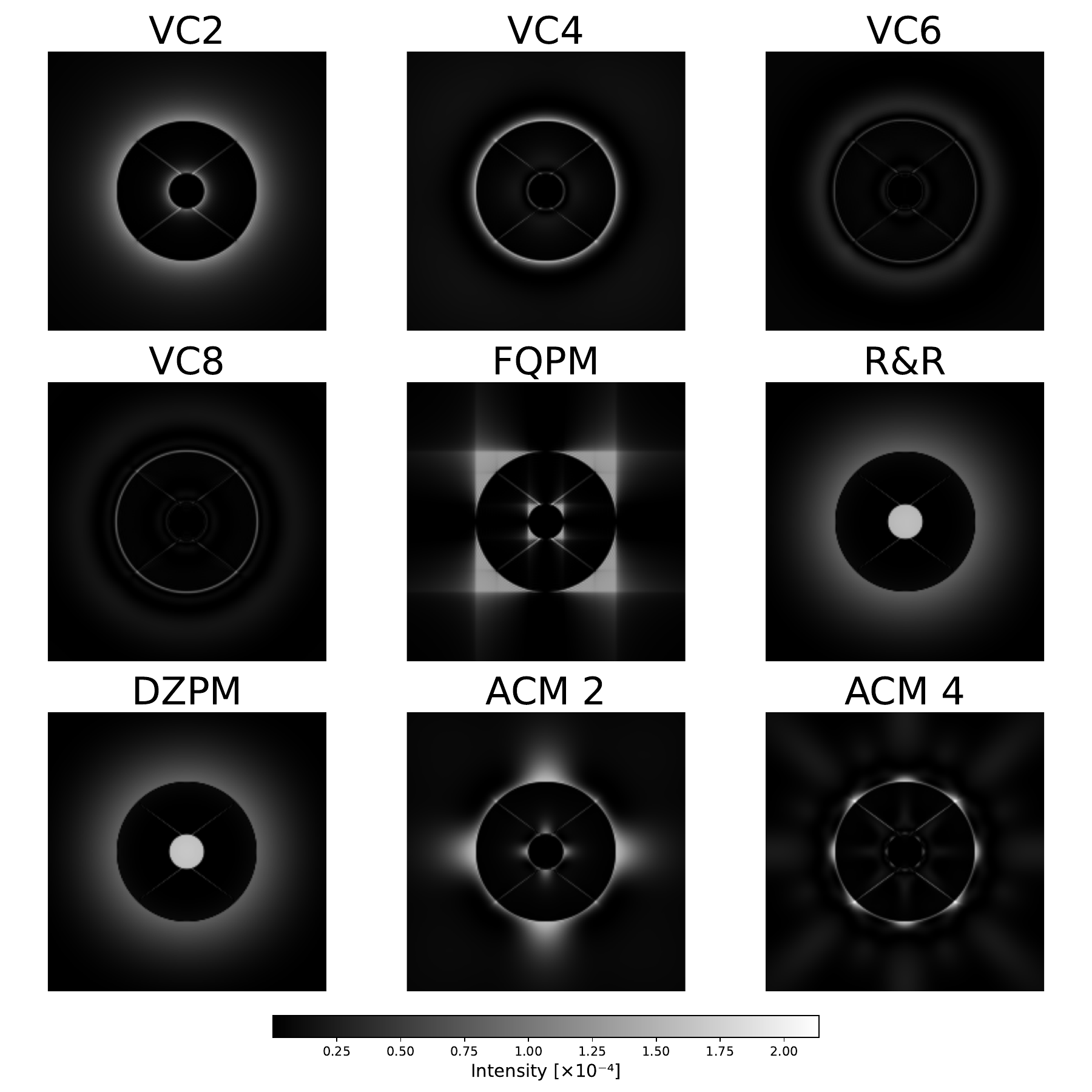}\end{tabular}
\end{center}
\caption 
{ \label{fig:lyot_plane}
Illustration of the light distribution in the LPP for nine commonly used FPMs. The entrance pupil includes a 25 $\%$ central obstruction and spider arms. The FPM features a spatial sampling of 10 pixels/($\lambda$/D) and an
8-bit grayscale phase resolution. Simulations assume monochromatic light and no wavefront errors.} 
\end{figure}

The light distribution at the lyot pupil plane provides valuable insight into the behavior of different phase masks under realistic conditions. Figure \ref{fig:lyot_plane} shows the light distribution at the LPP for nine commonly used phase masks discussed in this paper considering the presence of a central obstruction in the entrance pupil. The light distribution after FPM R$\&$R and DZPM differ noticeably from from those of the other designs. In these cases, most of the light leakage caused by the secondary obstruction is concentrated inside the secondary mirror region, allowing the Lyot stop obstruction to block the diffracted light more effectively without any oversizing. This behavior may explain the robust performance observed with the R$\&$R and DZPM.

\section{Mean contrast between 3 and 10 $\lambda$/D vs wavelength}
\label{app:mean_contrast}
\begin{figure}[H]
\begin{center}
\begin{tabular}{c}
\includegraphics[height=5.5cm]{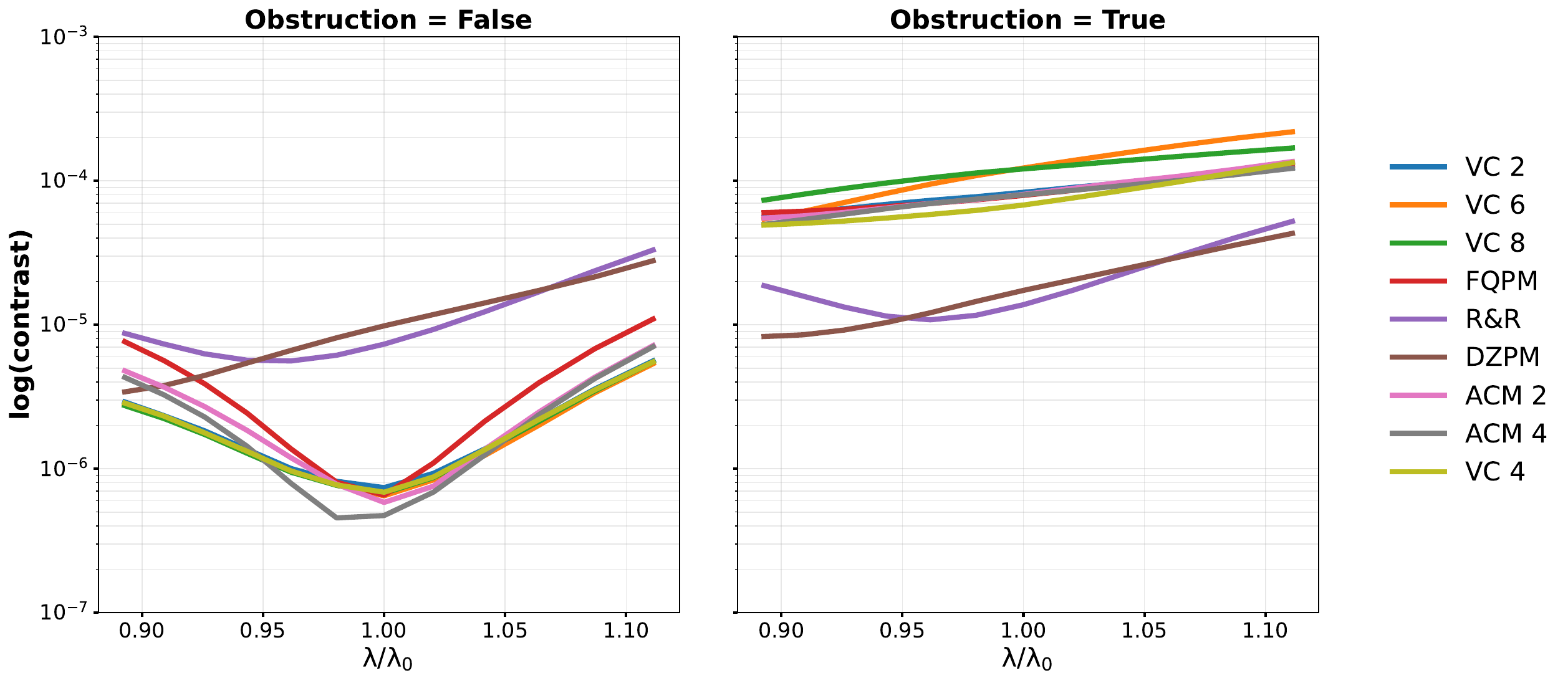}
\end{tabular}
\end{center}
\caption 
{ \label{fig:bb_crosssection_100}
Mean raw azimuthally averaged contrast in the SFP calculated between 3 and 10 $\lambda/D$ as a function of wavelength, ranging from $0.9\,\lambda_0$ to $1.1\,\lambda_0$.
    Nine commonly used phase masks are evaluated. The left panel shows results for an unobstructed entrance pupil, while the right panel shows results with a 25 $\%$ central obstruction and Lyot stop's secondary obstruction of $D_s'/D_s$ = 1. The FPM spatial sampling is set to 100 pixels/($\lambda$/D). Simulations assume no wavefront errors and 64-bit greyscale phase resolution.} 
\end{figure}

Figure \ref{fig:bb_crosssection_100} shows the mean contrast between 3 to 10 $\lambda$/D for different FPMs as a function of wavelength values, ranging from 0.9 $\lambda_{0}$ to 1.1 $\lambda_{0}$. This figure serves as a comparison to Figure \ref{fig:broadband_crosssection} but this time with an ideal FPM spatial sampling of 100 pixels/($\lambda$/D) and 64-bit resolution. The results indicate that the offset of the minimum contrast for the VC4 observed in Figure \ref{fig:broadband_crosssection} from the central wavelength arises from the limited spatial sampling of the FPMs.

\section{Residual energy: R$\&$R vs Dual zone}
\label{sec:app_res_ene}
\begin{figure}[H]
\begin{center}
\begin{tabular}{c}
\includegraphics[height=7cm]{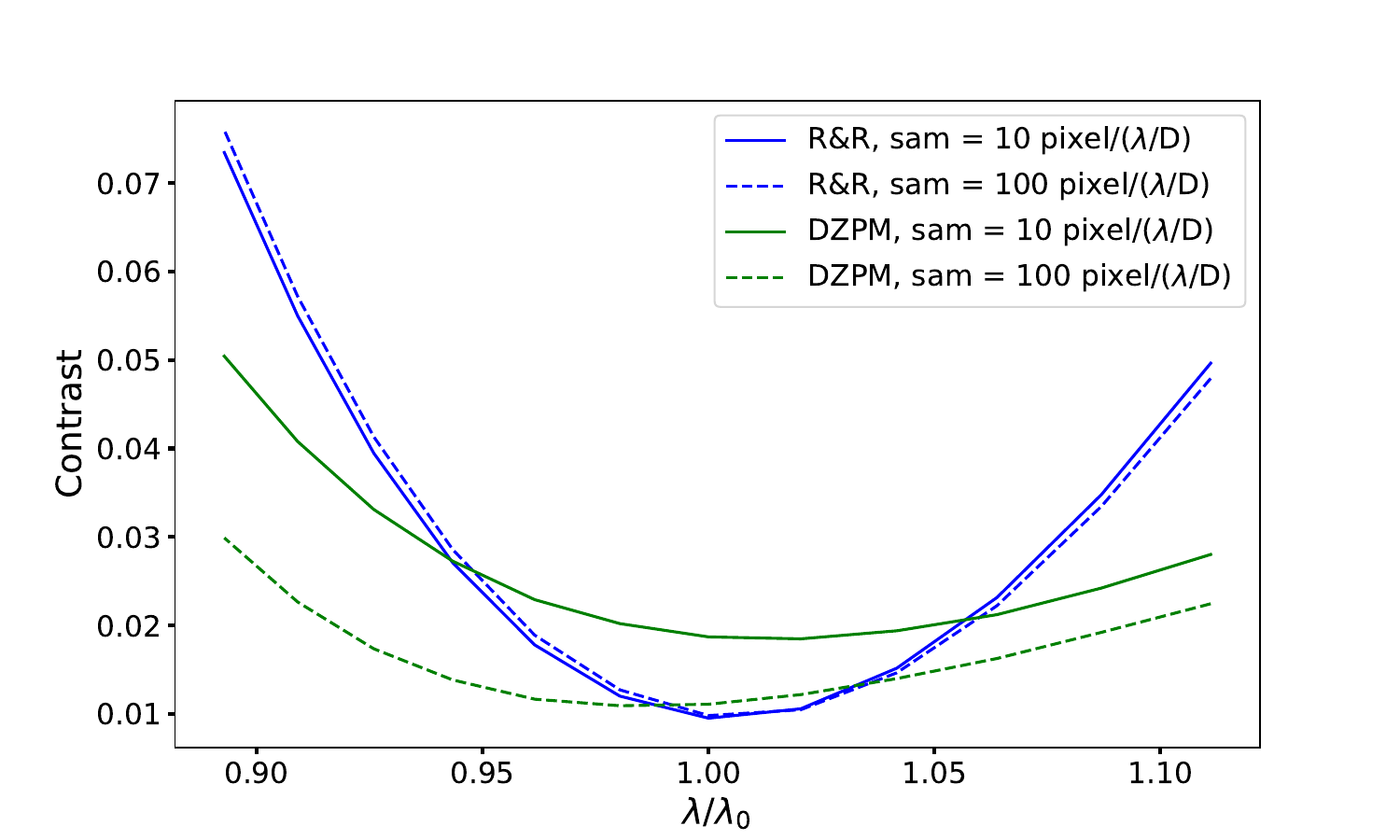}\end{tabular}
\end{center}
\caption 
{ \label{fig:resisual_ene}
Residual energy measured before LPP for the coronagraph system with R $\&$ R and Dual zone phase masks as a function of wavelength, ranging from $0.9\,\lambda_0$ to $1.1\,\lambda_0$. The entrance pupil is unobstructed. The FPM features a spatial sampling of 100 pixels/($\lambda$/D) and a
64-bit grayscale phase resolution. Simulations assume no wavefront errors.} 
\end{figure}

The energy ratio after and before the LS provides an additional metric for evaluating the performance of FPMs. In Figure \ref{fig:resisual_ene}, DZPM clearly presents higher robustness with broadband light compared to the R$\&$R that is less apparent in the mean contrast plots in Figures \ref{fig:broadband_crosssection} and \ref{fig:bb_crosssection_100}, the latter being generated with 100 pixels/($\lambda$/D) and 64-bit phase resolution.

\section*{Disclosures}
The authors declare that there are no financial interests, commercial affiliations, or other potential conflicts of interest that could have influenced the objectivity of this research or the writing of this paper.

\section* {Code, Data, and Materials Availability}
 The code and data used in this work are publicly available on GitHub: \url{https://github.com/LiurongLin/active_coronagraph}. The authors used Codex (OpenAI) to assist with refactoring research code used in this study. The tool was used to improve code organization and readability, without changing the scientific methods. Example prompts included: “Refactor this function for readability while preserving the original functionality.” All AI-generated suggestions were manually reviewed, tested, and validated by the authors.

\section* {Acknowledgments}
The authors used ChatGPT (OpenAI) to assist with English grammar and spelling, and clarity editing during manuscript preparation. Example prompt: “Please correct this paragraph for grammar without changing the original meaning.” All suggested edits were reviewed and approved by the authors, who take full responsibility for the final text.

The RACE-GO project has received funding from the Swiss State Secretariat for Education, Research and
Innovation (SERI), under the ERC replacement scheme following the discontinued participation of Switzerland
to Horizon Europe. Part of this work has been carried out within the framework of the National Centre of
Competence in Research PlanetS supported by the Swiss National Science Foundation under grants 51NF40
182901 and 51NF40 205606.

\bibliography{report}   
\bibliographystyle{spiejour}   

\vspace{2ex}\noindent\textbf{First Author} began her studies in the Astronomy program at Leiden Observatory in 2019 and later finished her Master's program in Astronomy and Data Science in early 2023. During her time in Leiden, she performed research in the exoplanetary atmosphere and planetary disk. In 2023, she was awarded a PhD candidate position in stellar coronagraphy instruments at the University of Bern. She is currently developing a fast SLM based coherent differential imaging framework that mitigates quasi-static and post-AO speckles to improve the contrast of exoplanet direct imaging. 

\vspace{2ex}\noindent After a master’s degree in astronomy at the University of Vienna, Ruben Tandon is currently studying for a PhD in astrophysics at the University of Bern, supervised by Prof. Jonas G. Kühn. As part of this project, he works on PLACID - the world’s first adaptive coronagraph - situated at the 4-m DAG telescope in eastern Turkey. Using the instrument for example for directly imaging exoplanets and binaries, he is responsible for the first on-sky results.

\vspace{2ex}\noindent After a Doctor degree of instrumentation for astronomy at the University of Paris-Saclay, Lucas Marquis is currently working as a Postdoc in astrophysics at the University of Bern (2024-2027), supervised by Prof. Jonas G. Kühn. As part of this project, he works on the first adaptive, programmable coronagraph PLACID for directly imaging exoplanets from the new 4-m DAG telescope in eastern Turkey (DAG) and is responsible for the implementation of breakthrough methods on the related laboratory's facilities.

\vspace{2ex}\noindent Following a PhD at EPFL, Lausanne, Jonas Kuhn obtained a fellowship at Caltech-NASA/JPL from 2012 to 2015, working on high-contrast imaging instruments for large telescopes. From 2015 to 2018, he took a PI position at ETH Zurich, pioneering the use of SLMs as active coronagraphs. Since 2019 he is the PI of the PLACID instrument for the 4-m DAG telescope, and since 2022, he is leading the ERC RACE-GO Project at the University of Bern.

\listoffigures
\listoftables

\end{spacing}
\end{document}